# HAADF-STEM Study of Mo/V Distributions in Mo-V-Te-Ta-O M1 Phases and Their Correlations with Surface Reactivity

Jungwon Woo[1], Albina Borisevich[2], Qian He [2] and Vadim V. Guliants*[1]

[1] Department of Biomedical, Chemical, and Environmental Engineering, University of Cincinnati, Cincinnati, OH 45221-0012, USA

[2] Center for Nanophase Materials Sciences, Oak Ridge National Laboratory, Oak Ridge, Tennessee 37831, USA

*Corresponding author: Vadim V. Guliants: Vadim.Guliants@UC.EDU

**Abstract**

The MoVTeTaO M1 phases were prepared by conventional hydrothermal (HT) and microwave-assisted HT synthesis methods (MW) employing two different Ta precursors, Ta ethoxide and a custom-made Ta oxalate complex. The profile intensity analysis of the HAADF-STEM image of M1 phases oriented along [hk0] directions from the surface to bulk region of HAADF-STEM images indicated that the chemical composition of surface *ab* planes is very similar to their composition in the bulk. The HAADF-STEM image analysis showed that synthesis methods have a significant impact on the Mo/V distribution in the MoVTeTaO M1 phases and their reactivity in propane ammoxidation. Enhanced acrylonitrile (ACN) yield and 1st order irreversible reaction rate constants for propane consumption ($k''_{ab}$), normalized to the estimated surface *ab* plane areas, correlated with increased V content in the proposed catalytic center (S2-S4-S4-S7-S7). These observations lend further support to the idea that multiple $VO_x$ sites present in the surface *ab* planes may be responsible for the activity and selectivity of the M1 phase in propane ammoxidation.

# 1. Introduction

The detailed atomic-scale information about the surface structure and chemical composition of crystalline mixed metal oxides that are currently investigated for the direct propane ammoxidation to acrylonitrile (ACN) is critical for understanding the fundamental surface structure-reactivity relationships and further improving these catalytic materials for this practical application. However, the characterization of topmost surface in such polycrystalline mixed metal oxides, as the M1 phase, remains a long-standing challenge in materials science of catalytic materials because the vast majority of spectroscopic and structural tools are bulk techniques [1-4]. In recent years, High-Angle Annular Dark-Field Scanning Transmission Electron Microscopy (HAADF-STEM) possessing sub-angstrom resolution emerged as a promising tool to study the structure and compositions of crystalline multicomponent metal oxides [5-7]. Pyrz et al. [8] analyzed the HAADF-STEM images of the M1 phase and determined the local chemical composition of the *ab* planes of this phase proposed as the location of the active centers for propane ammoxidation. However, the local chemical compositions of the *ab* planes of the MoVTeTa M1 phase determined in previous HAADF-STEM studies were essentially bulk compositions based on average intensities of entire atomic columns corresponding to a large number of *ab* planes [8,9].

Therefore, the first major objective of this study was to explore the potential of HAADF-STEM to analyze the composition of the topmost *ab* planes of the M1 phase. This profile HAADF-STEM study probed the intensities of atomic columns, which were entirely located within the surface *ab* planes by investigating the [hk0] crystal orientations of the M1 phase. Moreover, this profile HAADF-STEM technique enabled examining the variation of the *ab* plane composition from the topmost surface to the bulk. This composition analysis was accomplished by applying the same HAADF-STEM methodology that was used previously for the bulk characterization of the M1 phase oriented down the [001] crystal axis.

The second major objective of this study was to examine the location, concentration and catalytic role of Nb (and Ta) in propane ammoxidation to acrylonitrile. Although the presence of Nb is known to significantly improve the activity and selectivity of the M1 phase in propane (amm)oxidation [10], its location in the M1 phase could not be directly established by X-ray and neutron diffraction methods due to similar scattering properties of Nb and Mo centers. Instead, Pyrz et al. [11] investigated the Ta-substituted M1 phase by the HAADF-STEM and determined that Ta was located in so-called pentagonal bipyramidal site 9. It was further suggested that Nb being chemically similar to Ta is also located in site 9. Based upon this structural model of the M1 phase, Grasselli et al. [12-14] proposed a hypothetical pathway of propane ammoxidation over the *ab* planes of the M1 phase. According to this pathway, Nb cations, located in site 9 and surrounded by five Mo cations, spatially isolate adjacent V-containing active sites from one another thereby improving the selectivity of propane ammoxidation to acrylonitrile (Figure 1).

However, the M1-Ta catalyst prepared by a rapid slurry evaporation method [11] also revealed some unusual microstructural features which were expected to be detrimental for the selectivity to acrylonitrile according to the site isolation model [15]. For example, the HAADF-STEM images of this catalyst indicated much lower Ta occupancy in site 9 than that in the structural model based on a combination of synchrotron X-ray powder diffraction (S-XPD) and neutron powder diffraction (NPD) methods [16]. Moreover, the Ta segregation from the surface to the bulk was observed in a previous HAADF-STEM study of the M1-Ta catalyst [11]. However, the Ta, V, and Mo locations and partial site occupancies in hydrothermally synthesized

M1 phases have not been yet investigated despite the fact that hydrothermal synthesis offers improved control over the nucleation and growth of polycrystalline mixed metal oxides [17-19].

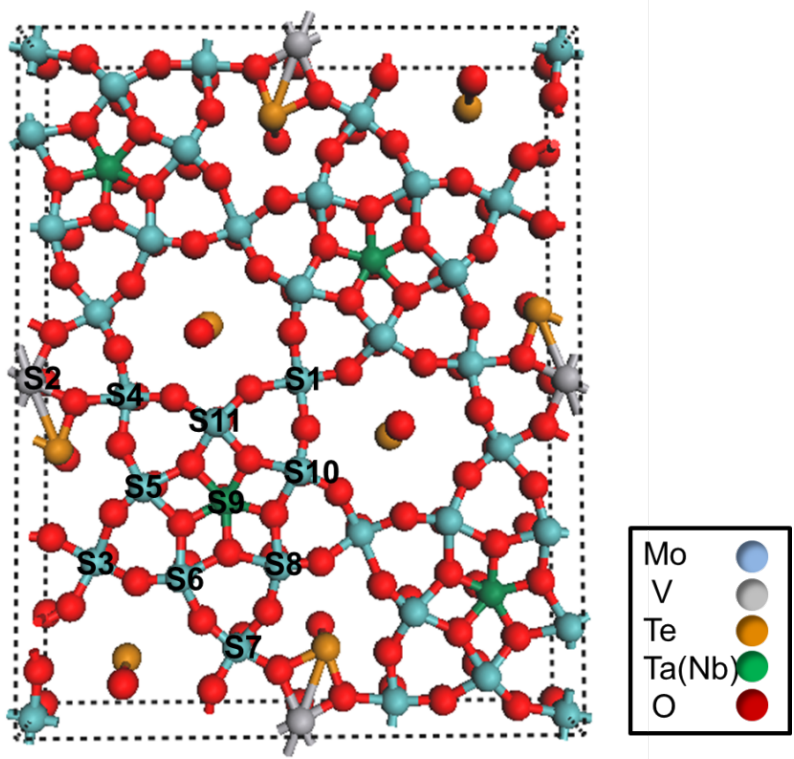

**Figure 1.** The crystal structure of MoVTeTa(Nb)O M1 phases. Nb is shown in the same S9 site [24]

In this study, the synthesis of well-defined MoVTeTaO M1 phase catalysts was explored employing approaches aimed at enhancing the Ta occupancy of site 9 and minimizing its surface segregation. Well-defined MoVTeTaO M1 phase catalysts were prepared by conventional hydrothermal (HT) and microwave-assisted HT synthesis methods (MW) employing two different Ta precursors, Ta ethoxide and a custom-made Ta oxalate complex (denoted as "P" below). The HAADF-STEM images of these M1 phase catalysts were obtained at Oak Ridge National Laboratory (ORNL). The atomic column intensities of the HAADF-STEM images of M1-Ta catalysts oriented along the [hk0] and [001] directions were collected and analyzed to determine whether surface *ab* planes had similar chemical composition to that of bulk *ab* planes and to provide the atomic scale metal distributions in the *ab* planes of the M1 phase. The analysis of HAADF-STEM images collected along the [001] direction provided the estimates of Mo, V, and Ta concentrations in the proposed catalytic centers of the MoVTeTaO M1 phases. The correlations between Mo, V, and Ta distributions in the proposed catalytic centers present in the *ab* planes and catalytic behavior of these model M1 phase catalysts are discussed below.

## 2. Experiment

*2.1 M1 phase catalyst preparation*

The M1 phase catalysts with synthesis molar ratios of Mo:V:Te:Ta=1:0.31:0.22:0.09, 0.12, and 0.15 were prepared by hydrothermal synthesis (HT) as previously reported [5, 23]. Ammonium paramolybdate (Alfa Aesar, 81-83% as $MoO_3$, 5.35 g) and telluric acid (Fluka, 99%, 1.16 g) were dissolved under stirring in 20 ml of distilled water for 30 min. A second solution was prepared by dissolving vanadyl (IV) sulfate (Alfa Aesar, 99.9%, 2.37 g) in 10 ml of distilled water and then a third solution was prepared by dissolving tantalum(V) ethoxide (99%, Alfa Aesar) or tantalum oxalate complex [20] in 10 ml of distilled water. The Ta oxalate complex was used as an alternative Ta precursor in the synthesis of MoVTeTaO M1 phase catalysts, which is more hydrolytically stable than the conventional Ta ethoxide source. The Ta oxalate complex was synthesized starting with $Ta_2O_5$ using the basic flux method. $Ta_2O_5$ was fused with KOH ($Ta_2O_5$:KOH = 1:10) in an alumina crucible at 723K for 5 h. The solid product was dissolved in distilled water, and the resulting solution was filtered. Glacial acetic acid was added to the transparent solution until a pH value below 3. The white precipitate was filtered, washed thoroughly to remove the remaining acetic acid, and dissolved in oxalic acid solution (Ta:oxalic acid = 1:20) at 333K. The Ta content in the Ta oxalate complex was determined by the ICP analysis and further checked by thermogravimetric analysis (TGA) in air. These characterization results were in good agreement with those reported in the literature [20]. The second solution for the M1 phase synthesis containing vanadium was added to the first solution of molybdenum and tellurium sources, and the resulting solution was stirred for 5 min. The third solution containing a tantalum source was finally added and the resulting yellow-green slurry was stirred for 10 min. The slurry was transferred into the Teflon inner tube of a stainless steel autoclave. The autoclave was sealed and heated at 448K for 48 h. After the hydrothermal reaction, the dark blue powder obtained was filtered, washed with distilled water (200 ml) and dried overnight at 353K. The dry catalyst precursors from HT method were calcined under ultra-high purity nitrogen flow (50 ml/min) at 873 for 2 h prior to catalytic studies.

In the case of the microwave-assisted synthesis (MW), the same slurry was prepared and transferred to a 100 ml capacity cylindrical Teflon vessel, sealed and then kept at 448K for 2 h in a microwave-accelerated Mars 5 reaction system (CEM, USA) operated at maximum power (400 W) [18]. After the microwave treatment, the dark blue powder was filtered, washed with distilled water (200 ml), and dried overnight at 353K. The dry catalyst precursor was calcined under ultra-high purity nitrogen flow (50 ml/min) at 873 for 2 h prior to catalytic studies.

In order to obtain pure M1 phases, the M2 phase impurity initially present in all calcined as-synthesized catalysts was selectively removed by a hydrogen peroxide treatment. The dissolution of the M2 phase was carried out by stirring the calcined as-synthesized catalysts in an aqueous 30% hydrogen peroxide solution at room temperature for 3 h [21]. The resulting suspension was filtered, washed with distilled water (200 ml), and dried overnight at 353K.

*2.2 Powder XRD characterization*

After the calcination at 873K for 2 h, the catalysts were thoroughly ground with a mortar and pestle for 10 min. Powder X-ray diffraction was recorded using a Siemens D500 diffractometer with Cu K$\alpha$ radiation (tube voltage: 45 kV, tube current: 40 mA).

*2.3 Scanning electron microscopy (SEM) and energy dispersive spectroscopy (EDS)*

The SEM and EDS characterization was conducted in the Advanced Materials Characterization Center (AMCC) at the University of Cincinnati. SEM (FEI/Philips XL 30 FEG ESEM) was equipped with the energy dispersive X-ray analyzer from EDAX and had a resolution of 3.5 nm at 30 kV.

*2.4 Specimen preparation for STEM*

The specimens employed in STEM studies were prepared as follows. The MoVTeNb(Ta)O M1 phase catalysts prepared by different synthesis approaches were thoroughly ground by a mortar and pestle for 5 min. The ground M1 samples were embedded into a resin and sectioned by a microtome into ca. 50 nm slices [22] in order to study the [001] zone axis orientation of the M1 phase. These specimens were then deposited onto a lacey-carbon supported Cu grid.

High-resolution STEM imaging was performed using an FEI-Titan 60/300 microscope equipped with a CEOS aberration corrector on the 300kV illumination system in the Advanced Microscopy Division of the Oak Ridge National Laboratory (ORNL). Z-contrast images in this work were acquired at 300 kV with a probe size of ~0.6 Å, a convergence angle of 30 mrad, and inner collection angle of 60 mrad. The acquisition time used to collect these images was 6 secs/1024*1024 pixels. All specimens were plasma-cleaned using Fischione plasma cleaner by electron beam irradiation to prevent specimen contamination during the STEM operation.

*2.5 Propane Ammoxidation Reaction*

The catalytic behavior of the MoVTeTa(Nb) M1 phase catalysts prepared by different synthesis methods was tested in propane ammoxidation using a fixed bed micro-reactor equipped with an on-line GC under steady-state conditions at atmospheric pressure and 380 – 460 °C. The calcined catalysts were ground with a mortar and pestle for 5 min, diluted with quartz sand, loaded into the micro-reactor, and then heated to a desired temperature in flowing feed gas. The feed was composed of $C_3H_8$:$NH_3$:$O_2$:He in the molar ratio of 5.7:8.7:17.1:68.4 (the total flow rate of 26.3 mL•$min^{-1}$). The reactants and products were analyzed by an on-line GC system (Shimadzu 14 A) equipped with flame ionization and thermal conductivity detectors. The total carbon balances agreed within ±5%.

*2.6 STEM image analysis*

In HAADF-STEM image analysis, the center-of-mass fit determined the positions and intensities of atomic columns. In order to interpret the intensity quantitatively in terms of elemental occupancies, the following assumptions were made: (1) the thickness of each unit cell was constant; (2) the background intensity was constant throughout each unit cell; and (3) the atomic column intensity was proportional to the square of the atomic number following the Rutherford scattering relationship [23]. Based on these assumptions, the intensities of atomic columns were computed using Digital Microscopy (DM) scripts. The obtained atomic column intensities were normalized to the average intensity of the 100% Mo sites S5, S6, S8, S10, and S11 [24].

## 3. Results and Discussion

*3.1 Preparation of pure M1 phases*

Figure 1 of Supporting Information shows the XRD patterns of the calcined MoVTeTa(Nb)O M1 catalysts prepared by different synthesis methods after the $H_2O_2$ treatment, namely hydrothermal (HT) and microwave-assisted synthesis (MW) using two different Ta precursors, the Ta oxalate complex and Ta ethoxide. M1-Ta HT synthesized using the Ta oxalate complex is labeled below as M1-Ta HT(P). The M1-Ta HT phases made using different Ta concentrations during synthesis are denoted as M1-Ta HT (0.09), M1-Ta (0.12), and M1-Ta (0.15), respectively, where the numbers in parentheses indicate the synthesis Ta/Mo ratios. All catalysts after the $H_2O_2$ treatment showed very similar diffraction patterns (Figure 1 of Supporting Information), i.e., peaks at $2\Theta$ = 7.8, 8.9, 22.1, 27.2, and 45°, which are indicative of pure M1 phase (PDF 01-073-7574). These XRD patterns confirmed that the different synthesis methods employed in this study, i.e., HT, MW, and HT(P), result in well-defined MoVTeTa(Nb)O M1 phase catalysts besides the slurry evaporation (SE) method used previously [12,16]. It is noteworthy to mention that M1-Ta HT(P), made using the Ta oxalate complex as an alternative Ta source, was synthesized for the first time in this study. The Ta oxalate complex was employed instead of Ta ethoxide, conventionally used in the synthesis of MoVTeTaO catalysts, because it was expected to result in enhanced Ta incorporation in site 9 due to its greater hydrolytic stability. The resulting M1-Ta HT(P) also contained pure M1 phase similar to HT and MW methods where the Ta ethoxide was employed as the Ta source. This result confirmed that the Ta oxalate complex is an alternative Ta source for the synthesis of MoVTeTaO M1 phase catalysts. Freshly calcined as-synthesized catalysts of this study contained some M2 phase impurity which was removed by the $H_2O_2$ treatment. It was observed that M1-Ta (0.09) and (0.15) contained considerable amounts of the M2 phase, while the M1-Ta (0.12) contained an almost pure M1 phase after the calcination (XRD patterns of catalysts before the $H_2O_2$ treatment are not shown). These results suggested the importance of Ta concentration during synthesis as one of the key parameters determining the final phase composition of the MoVTeTaO catalysts.

*3.2 SEM images and elemental composition of MoVTeTaO M1 phase catalysts*

The representative SEM images of the MoVTeTaO M1 catalysts are shown in Figure 2 of Supporting Information. All SEM images showed typical rod-like particle morphology of M1 phases regardless of the synthesis methods employed. It is important to note that the morphology of the MoVTeTaO M1 phases of this study was different from those of the previously reported MoVTeTa(Nb)O catalysts prepared by slurry evaporation (SE) (Figure 2f of Supporting Information) [16,25]. This result is not surprising because the SE method is known to result in rather poor control over nucleation and growth of the desired M1 phase, and often lead to the formation of M2 and other impurity phases [18]. As shown above, the SEM images of hydrothermal MoVTeTaO M1 phases, i.e., HT (0.09, 0.12, 0.15), MW, and HT(P), showed a much better defined crystal morphology and typical rod-like crystal habit of the M1 phase (Figure 2 of Supporting Information). These observations indicated that the synthesis method has a profound impact on the crystal morphology of the MoVTeTaO M1 phase catalysts. Moreover, they suggested that HT and MW synthesis methods yield well-defined M1 phases that are better suited for establishing accurate relationships between the metal distributions in the *ab* planes of the M1 phase and its catalytic performance in propane ammoxidation.

*3.3 Bulk characteristics of MoVTeTaO M1 catalysts*

All MoVTeTaO M1 catalysts were characterized by SEM/EDS with respect to their bulk chemical compositions and BET surface areas (Table 1). The bulk elemental composition of all catalysts differed slightly depending on the synthesis methods. It is generally observed that the bulk Ta/Mo molar ratio of all catalysts determined by SEM/EDS is much higher than that in the synthesis gel. The results of previous studies [11,16,24] indicated that Ta is exclusively located in site 9 (S9). If we assume that Ta fully occupies S9 and is not present anywhere else in the M1 lattice, while the remaining sites are occupied by 70% Mo and 30% V, then the Ta/Mo ratio would be ca. 0.14. Since the observed Ta/Mo ratios (Table 1) are markedly higher than 0.14, the higher molar ratio of Ta/Mo suggested that the excess Ta is likely present outside the M1 phase lattice. This observation is supported by the results of DeSanto et al. [16], who detected a $Ta_2O_5$ impurity in the M1-Ta sample prepared by slurry evaporation in a combined synchrotron (S-XPD) and neutron power diffraction (NPD) study of this phase. Moreover, this observation also suggested that not all Mo was incorporated into the M1 phase during synthesis.

**Table 1**. Bulk characteristics of MoVTeTaO catalysts prepared by different synthesis methods.

| Catalysts | Preparative[a] | SEM/EDS[b] | Surface area $(m^2/g)$[c] | |
|---|---|---|---|---|
| | Mo/V/Te/Ta | Mo/V/Te/Ta | Before $H_2O_2$ treatment | After $H_2O_2$ treatment |
| M1-MoVTeTa HT (0.09) | 1.00/0.31/0.22/0.09 | 1/0.28/0.08/0.28 | 22.4 | 37.2 |
| M1-MoVTeTa HT (0.12) | 1.00/0.31/0.22/0.12 | 1/0.32/0.17/0.33 | 4.3 | 14.1 |
| M1-MoVTeTa HT (0.15) | 1.00/0.31/0.22/0.15 | 1/0.30/0.10/0.34 | 4.9 | 12.3 |
| M1-MoVTeTa MW | 1.00/0.31/0.22/0.12 | 1/0.27/0.17/0.21 | 3.1 | 17.7 |
| M1-MoVTeTa HT(P) | 1.00/0.31/0.22/0.12 | 1/0.22/0.10/0.20 | 2.1 | 17.5 |

[a] Synthesis composition of the slurry; [b] Determined by EDS; [c] Determined by the BET method; MW (Microwave-assisted hydrothermal synthesis); HT (Hydrothermal synthesis); P (made using Ta oxalate as Ta precursor)

It is well known that the $H_2O_2$ treatment selectively removes the M2 phase, which contains more Te and less Nb (or Ta) as compared to the M1 phase [25]. The $H_2O_2$ treatment effect is confirmed by the decrease of the Te concentration and increase of the Nb (or Ta) concentration after the $H_2O_2$ treatment (Table 1). Although the M1-Ta HT (0.15) made at the highest synthesis Ta concentration also had the highest Ta concentration in the bulk among all M1-Ta catalysts, there appears to be no strong correlation between the Ta concentration in synthesis and bulk determined by SEM/EDS shown in Table 1. It is observed that M1-Ta HT (0.09) has a much higher BET surface area as compared to other MoVTeTaO M1 phase catalysts.

*3.4 HAADF-STEM imaging of MoVTeTaO catalysts*

*3.4.1 Ta location in ab planes of MoVTeTaO M1 phase catalysts*

The results of HAADF-STEM imaging of the slurry evaporation MoVTeTa(Nb)O M1 phase catalysts reported by Pyrz et al. [8,9,11] were in good agreement with the full structural model of the M1 phase based on the S-XPD and NPD refinement [16]. Therefore, we further applied the HAADF-STEM to studies of the M1-Ta catalysts given the atomic-scale imaging capability and elemental composition information offered by this method. Firstly, the Ta segregation observed in these previous studies was addressed by employing catalysts made by the HT and MW methods, which allow better control over the formation of desired M1 phase than the slurry evaporation method used previously [11]. Furthermore, we employed specimens of constant thickness (~ 50 nm) for HAADF-STEM imaging made by sectioning the embedded samples to eliminate the impact of thickness variation on the atomic column intensities.

Representative HAADF-STEM images of M1-Ta HT(0.12) which were recently reported by our group [24] were shown in Figure 3 of Supporting Information. The HAADF-STEM images showed that the pentagonal sites (S9), located at the center of circles visible in Figure 3a of Supporting Information, were much brighter than other metal sites. The significant brightness contrast between S9 and all other metal lattice sites, including those occupied by Mo, clearly indicated the location of heavy Ta atoms in S9. There were no appreciable contrast differences between S9 column located in the middle of the M1 crystal (Figure 3b of Supporting Information) and its surface region (Figure 3c of Supporting Information). These findings confirmed that Ta is uniformly distributed throughout the M1-Ta HT catalyst. Uniform Ta distribution in M1-Ta HT suggested that the Ta segregation observed in the previous M1-Ta SE is likely due to mass-transfer limitations during rapid slurry evaporation synthesis [16].

Density functional theory (DFT) was further employed to probe the energies of Ta and Nb cations located in several crystallographic sites of the M1 phase [24]. These spin-polarized periodic DFT calculations were performed in the generalized gradient approximation (GGA-PBE) [26] using the Vienna Ab initio Simulation Package (VASP) [27-30]. Three truncated *ab* planes of the M1 phase structure were employed as cluster models for the DFT calculation. The Ta located at site 9 was interchanged one by one with other metal sites present in the M1 phase and the energies of cluster models resulting from the Ta interchange were calculated from DFT by applying statistical thermodynamics considering both the internal energy and entropy contributions to the free energy of the Ta distribution for the *ab* planes. The same computational approaches were applied to Nb in the cluster models of M1 phase catalyst. The predicted probabilities of finding Ta and Nb in S9, 10, and 11 determined by DFT calculations combined with methods of statistical thermodynamics are shown in Figure 3 of Supporting Information. The probability of finding Ta and Nb in S9 is predicted to be nearly 100% confirming that Ta (and Nb) is predominantly located in S9 in good agreement with the results of HAADF-STEM image analysis of the M1-Ta phase.

A comparison of relative intensities of metal lattice sites for the MoVTeTa(Nb)O M1 catalysts prepared by different synthesis methods is shown in Figure 4 of Supporting Information. The relative intensity for M1-Nb HT is also shown for comparison between the M1-Nb and M1-Ta systems. The normalization was done in reference to the average intensities of S5, S6, S8, S10, and S11 which were assumed to be 100% Mo (Z=42). This assumption was based on the results of earlier X-ray, neutron diffraction, and HAADF-STEM studies [11,31,32], which reported S5, S6, S8, S10, and S11 to be 100% Mo sites. S5, S6, S8, S10, and S11 indeed showed constant intensities throughout all unit cells examined in this study. The higher intensities (greater than 1) of S9 atomic columns relative to 100% Mo sites suggested the presence of a heavier atom, i.e., Ta (Z=73), while the relative intensities less than 1 for S1, S2, S3, S4 and S7 indicated the presence of a lighter metal atom, i.e., V (Z=23), in agreement with results of the previous synchrotron X-ray diffraction and HAADF-STEM studies [11,16]. Therefore, the analysis of normalized intensities of atomic columns in the HAADF STEM images of M1-Ta catalysts employed in this study strongly indicated that Ta is predominantly located in site 9. It is noteworthy to observe significant variation of relative intensity of Ta in S9 as well as V in S1, S2, S3, S4, and S7 among M1-Ta catalysts shown in Figure 4 of Supporting Information. These findings suggested that different synthesis approaches have a marked impact on the Ta and V occupancies in the *ab* planes of M1 phase, which is discussed in greater detail below.

*3.4.2 Intensity line profile of HAADF-STEM images of M1-Ta HT (0.12) oriented down [hk0]*

A line profile collects scattering intensity along a line in an image and displays the intensity data in a line spot using Gatan Microscopy (GM) software. The line profile analysis of HAADF-STEM images has been previously employed as an important method for determining superficial and interfacial structures in various materials. For example, Liu et al. [33] employed the line profile STEM analysis to study the surface channeling phenomenon in a ZnO nanostructure, where the ZnO surface had an abnormally enhanced intensity despite the presence of structurally perfect surface because the diffracted electron beam was trapped inside the ZnO crystal. They observed the surface channeling effect from the line profile across the surface and interfaces of the HAADF-STEM image where the enhanced intensity of a row of atoms was collected from the near-surface region of the ZnO structure.

The line profile analysis of HAADF-STEM images was also employed to examine the $PbTiO_3$ film growth on a $SrTiO_3$ single crystal [34]. Significant line profile intensity differences between $PbTiO_3$ and $SrTiO_3$ in HAADF-STEM images of epitaxial $PbTiO_3$ perovskite grown on a $SrTiO_3$ single crystal enabled distinguishing the formation of a nanoscale $PbTiO_3$ film on a $SrTiO_3$ crystal. In order to study the effect of surface strain relaxation on HAADF images, three quantum wells (QW) of $In_xGa_{1-x}As$ with increasing In content grown by molecular beam epitaxy along the GaAs [001] direction were prepared. Due to the difference of In and Ga atomic numbers, three InGaAs QWs were distinguished by intensity contrasts due to the In concentration variation observed in the line profile of HAADF images [35]. In yet another study, the projection along the [001] crystal direction of AlAs/GaAs system grown on the top of a GaAs substrate was analyzed in high-resolution HAADF images obtained from super STEM [36]. The HAADF intensity profiles of three GaAs dumbbells were used to estimate the average Al and Ga content due to the Z-contrast nature of HAADF images, i.e., Al (Z = 13) and Ga (Z = 31). The line profile of HAADF-STEM images was also used to determine atomic scale structures of Pt monolayer electro-catalysts in combination with other characterization techniques, such as energy-dispersive X-ray spectrometry (EDS), electron energy-loss spectroscopy (EELS), and in situ extended X-ray absorption fine structure (EXAFS) [37].

Accordingly, we further employed the line profile HAADF-STEM analysis in this study as a tool to probe the surface and bulk of MoVTeTaO M1 phase catalysts given its demonstrated capabilities for the analysis of surfaces and interfaces of various solid-state materials with atomic-scale elemental resolution. The distinct advantage of the line profile HAADF-STEM analysis as applied to the M1 phase is rooted in its layered nature built by the stacking of *ab* planes along the *c* direction (Figure 5 of Supporting Information). Therefore, it is possible to probe the intensity profiles of the *ab* planes layer by layer from the topmost surface to bulk along the [hk0] directions due to layered structure of the M1 phase.

The HAADF-STEM images of the M1-Ta HT(0.12) particle oriented along the [hk0] direction and corresponding intensities from the line profile of 10 *ab* planes from the topmost surface towards the bulk are displayed in Figure 2. Figure 2a shows the typical rod-like M1-Ta HT particle viewed down the <hk0> direction. The line profile intensities of 10 *ab* planes (corresponding to a span of ca. 4 nm) were recorded along [hk0] (Figure 2c). All line intensity profiles were integrated over a width of 0.05 nm. The following two assumptions were made during this HAADF-STEM profile analysis:

(1) The absence of thickness variation from the collected area, and
(2) The validity of the Rutherford scattering relationship for observed intensities [23].

It was not possible to determine the specific *h* and *k* values for the [hk0] orientations in this STEM study because of the structural complexity of the M1 phase. Therefore, detailed chemical analysis of each atomic column and *ab* layer was not possible because the nature of metal cations in each atomic column could not be established due to the lack of knowledge of specific *h* and *k* values. Nevertheless, the absolute intensities of the structurally related atomic columns in different *ab* planes could be examined from the topmost surface to the bulk in a line profile scan. Therefore, the intensities of atomic columns in three topmost layers, corresponding to the surface region, were compared to those of seven other layers, considered representative of the bulk. According to Figure 2c, the line profile intensity (obtained by subtracting background intensity from absolute intensity) of each *ab* plane appears to be fairly constant throughout the first 10 layers. Therefore, these observations provide evidence that the chemical composition of the surface region of the M1 phase represented by the topmost 3 layers and the bulk represented by other 7 layers are similar. Slightly lower line profile intensity was observed near the surface (Figure 2c), which may be explained by the slight thickness variation in the [hk0] direction due to the presence of steps on the topmost surface. This conclusion is supported by the results of a recent HAADF-STEM study by Blom et al. [39] where they also found some thickness variation near the surface due to so-called "terracing" of the surface *ab* planes. The line profile intensity of the bulk M1 phase represented by the other 7 layers in Figure 2 clearly indicated that there was no significant intensity variation with depth. Therefore, these results indicated that the chemical composition of the surface and bulk *ab* planes is similar. In addition, there was no evidence related with disorder associated with the difference in the top surface as compared to the bulk from the HAADF-STEM simulation study of MoVTeNbO [40]. These highly important results indicated that the HAADF-STEM images of *ab* planes oriented along the [001] direction can provide the metal site occupancies in the surface *ab* planes due to the similarities of the surface and bulk compositions found in this profile HAADF-STEM study.

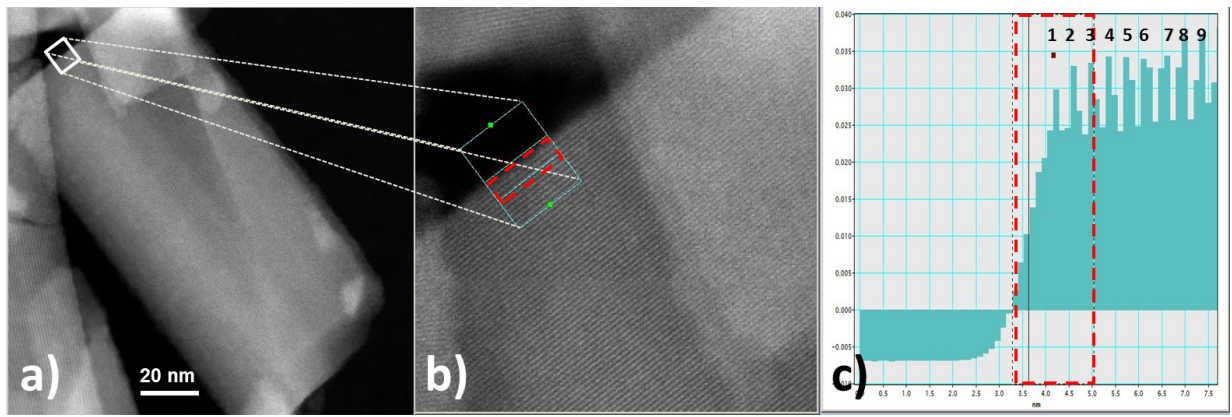

**Figure 2.** HAADF-STEM images of MoVTeTaO HT (0.12) M1 phase viewed along the [hk0] direction: a) low magnification image; b) magnified region, and c) intensity profile of top 10 layers along the *c* axis.

3.4.3 *Dependence of Mo/V distribution in ab planes on M1 phase synthesis methodology*

The Mo, V, and Ta occupancies in 11 crystallographic sites of the MoVTeTaO M1 phase prepared by different synthesis methods, i.e., HT, MW, and HT(P), were calculated based on the systematic analysis of atomic column intensities in the HAADF-STEM images viewed along the [001] direction (Table 1 of Supporting Information). The metal occupancies reported in Table 1 of Supporting Information were calculated based on the two assumptions stated in section 2.6. Three different M1 phase particles for each catalyst were chosen for analysis and two images for each particle were examined to determine the metal column intensities. The atomic column intensities in four unit cells in each image were thoroughly analyzed. The collected data for each catalyst were statistically analyzed and the atomic column intensities for each crystallographic site in all catalysts varied less than 10 % for the same site in different unit cells in the same M1 particle and different M1 phase particles. This finding indicates that the metal occupancies presented in present study are representative for the entire catalysts as a bulk material. The chemical composition of the M1-Ta prepared by slurry evaporation reported in a previous study [18] is also shown for comparison [16]. In a previous S-XPD study [16], sites 5, 6, 8, 10, 11 of the MoVTeTaO M1 catalyst prepared by slurry evaporation were reported to have full Mo occupancy. The Mo occupancy of sites 5, 6, 8, 10, 11 in all M1-Ta phases prepared by different synthesis methods (Table 1 of Supporting Information) is close to 100% which agrees well with results of the previous study [16]. However, significant variation of Mo and V occupancies in the linking sites 1, 2, 3, 4, and 7 was observed along with the Ta occupancy variation in site 9 for the M1 phases made by different synthesis methods. It should be noted that sites 1 and 4 were found to be mixed Mo/V sites in M1-Ta catalysts of this study, while they were fully occupied by Mo in the M1-Ta SE employed in the previous study [16]. This observation is very important because site 4 is an integral part of the hypothetical catalytic center (S2-S4-S4-S7-S7) for propane ammoxidation proposed by Grasselli et al. [12-14]. The impact of the V partial occupancy in site 4 on the catalytic performance in propane ammoxidation is discussed in the next section. Lastly, the variation of metal occupancies observed in M1 catalysts suggested that the partial occupancies of constituent metal species are strongly influenced by the choice of synthesis methods.

As mentioned above, Ta oxalate was employed as an alternative Ta source because of its greater solubility and hydrolytic stability as compared to Ta ethoxide. Therefore, the use of Ta oxalate was expected to enhance the incorporation of Ta in the M1 lattice as compared to Ta ethoxide. However, the M1-Ta HT(P) made from Ta oxalate resulted in the lowest Ta occupancy at S9 of the M1 phase (Table 1 of Supporting Information). This low Ta occupancy of S9 in M1-Ta HT(P) may be explained by the high stability of Ta oxalate complexes under conditions of M1-Ta HT(P) synthesis. This hypothesis is supported by the study of oxalate dissociation from a number of Ta hydroxyoxalate complexes observed in solution by Babko et al. [41]. They reported these dissociation constants to be in the range of ~ $8.0 \times 10^{-12}$ to $3.0 \times 10^{-19}$. Therefore, the Ta oxalate precursor employed in the M1-Ta HT(P) synthesis is expected to provide fewer free hydroxylated Ta species as compared to Ta ethoxide that are incorporated into the M1 phase lattice during synthesis.

*3.4.4 Correlations between Mo/V distributions and Ta content in synthesis*

The Mo, V, and Ta sites occupancies of the MoVTeTaO M1 phase catalysts prepared by the hydrothermal method using different synthesis Ta concentrations were determined by the systematic analysis of atomic column intensities from the HAADF-STEM images (Table 2 of Supporting Information).

This analysis revealed that sites 5, 6, 8, 10, and 11 are occupied almost entirely by Mo in M1-Ta HT (0.09), (0.12), and (0.15). This result is consistent with previous findings based on S-XPD [16] and the HAADF-STEM image analysis done by Pyrz et al. [11] as well as our results for the MoVTeTaO M1 phase catalysts prepared by different synthesis methods presented above (Table 1 of Supporting Information). Significant variation in Mo and V content was observed for sites 1, 2, 3, 4, and 7 similar to that observed for the MoVTeTaO M1 phase catalysts prepared by different synthesis methods (Table 1 of Supporting Information). In this study, we aimed to enhance Ta occupancy at S9 by increasing the Ta concentration in the synthesis gel. However, it was found that the Ta occupancy in S9 did not correlate with synthesis Ta concentrations. For example, M1-Ta (0.15) was expected to have the highest Ta occupancy in S9 among M1-Ta HT (0.09), (0.12), and (0.15), but the Ta S9 occupancy in M1-Ta (0.15) was slightly higher than that in M1-Ta (0.09), but lower than that in M1-Ta (0.12). These results suggested the importance of kinetic factors during the synthesis of M1-Ta catalysts that are insufficiently understood at present.

The Mo/V and Mo/Ta ratios for M1-Ta HT (0.09), (0.12), and (0.15) phases determined by different characterization techniques are listed in Table 3 of Supporting Information. The Mo/V ratios characterized by SEM/EDS, which is another bulk analysis technique, were in a good agreement with the synthesis Mo/V ratios. However, the Mo/V ratios from the HAADF-STEM were somewhat different from those in the synthesis gels. Furthermore, the Mo/Ta ratios determined from the HAADF-STEM images were significantly higher than both the synthesis and SEM/EDS ratios. These observations suggested the Ta presence outside the M1 phase lattice or a systematic STEM analysis error based on the assumption that n=2 for $Z^n$ that has much greater impact on the Ta concentration (Z=73) than those of lighter Mo (Z=42) or V (Z=23) atoms.

The SEM/EDS technique provided the overall elemental composition of MoVTeTaO M1 catalysts which might be complicated by the presence of some Ta impurity phase, e.g., $Ta_2O_5$, reported earlier [16]. However, the HAADF-STEM imaging directly probed the elemental composition of the MoVTeTaO M1. Much lower SEM/EDS Mo/Ta ratios than the synthesis ratios indicated that a significant fraction of Mo was not incorporated into M1 phase leading to a lower yield of M1 phase, e.g., ~ 20-30 at. % based on total Mo used in synthesis of all MoVTeTaO catalysts. Moreover, the results of the HAADF-STEM imaging indicated that Ta initially present in the synthesis gel is incorporated into the M1 lattice only partially and that site 9, in addition to Ta, has a significant Mo occupancy. Lastly, the metal site occupancies determined from the HAADF-STEM imaging were subsequently employed to build detailed surface molecular structure - catalytic reactivity relationships for propane ammoxidation to acrylonitrile over M1 phase catalysts. These fundamental structure-catalytic reactivity relationships are discussed in greater detail below.

*3.5 Propane ammoxidation over MoVTeTaO M1 catalysts*

*3.5.1 Catalytic performance of MoVTeTaO M1 phase*

M1-Ta HT(0.09), HT(0.12), HT(0.15), MW, and HT(P) were tested in propane ammoxidation reaction. The reaction data for propane ammoxidation for all catalysts was presented in Table 2. As reported earlier, propylene is the major product at low propane conversion, suggesting that propylene is the primary intermediate during propane ammoxidation, which is converted further to acrylonitrile, and then to unselective products, i.e., acetonitrile and $CO_x$, as propane conversion increases [4]. The propylene selectivity behavior found in this study is also in agreement with previous observations [4] where propylene selectivity was found to be higher at low propane conversion and decrease at higher conversion regardless of the Mo/V distribution in M1-Ta catalysts. The observed maximum selectivity to acrylonitrile under optimal reaction conditions decreased in the following order: M1-Ta HT(0.12) > M1-Ta HT(0.09) ≥ M1-Ta HT(0.15) ≥ M1-Ta MW = M1-Ta HT(P). The catalytic behavior reported in Table 2 is further discussed below on the basis of metal occupancies in *ab* planes of M1 phase catalysts.

**Table 2.** Reaction data for propane ammoxidation over M1-Ta HT(0.09), M1-Ta HT(0.12), M1-Ta HT(0.15), M1-Ta MW, and HT(P) catalysts.

| Catalysts | [a]T (°C) | Conversion (%) | Propylene (S) (%) | ACN (P) (%) | BET surface area (m$^2$/g) | [b]k"$_{BET}$ {m$^3$/(m$^2$ s)} | [c]$ab$ planes surface area (m$^2$/g) | [d]k"$_{ab}$ {m$^3$/(m$^2$ s)} |
|---|---|---|---|---|---|---|---|---|
| M1-Ta HT(0.09) | 420 | 28 | 10.5 | 65.3 | 37.2 | 4.62E-08 | 1.7 | 1.00E-06 |
| M1-Ta HT(0.12) | 420 | 14.7 | 18.4 | 76.9 | 14.1 | 5.87E-08 | 2.7 | 3.10E-07 |
| M1-Ta HT(0.15) | 420 | 10.6 | 27 | 70.5 | 12.3 | 4.74E-08 | 3.8 | 1.53E-07 |
| M1-Ta MW | 420 | 8.7 | 32.1 | 65.4 | 17.7 | 2.67E-08 | 2.6 | 1.81E-07 |
| M1-Ta HT(P) | 420 | 6.2 | 42.8 | 54.9 | 17.5 | 1.90E-08 | 3.3 | 9.95E-08 |

Reaction conditions: 0.2 g catalyst; C$_3$H$_8$:NH$_3$:O$_2$:He=5.7:8.6:17.1:68.6; the total flow rate of 26.3 mL•min$^{-1}$
[a]T is the reaction temperature; [b]k"$_{BET}$ is the reaction rate constant based on the BET surface area; [c]$ab$ plane surface areas were taken from Table 11 of Supporting Information; [d]k"$_{ab}$ is based on the $ab$ planes surface area

*3.5.2 Relationships between metal site occupancies in MoVTeTaO M1 phase catalysts and their catalytic performance*

*3.5.2.1 Relationships between Ta occupancy and Mo/V distribution in ab planes of M1 phase and its catalytic behavior*

The relationships between the Ta occupancy in site 9 of the M1 phase and the selectivity toward acrylonitrile were investigated on the basis of the site isolation model, defined as "the spatial separation of active sites from each other on the surface of a heterogeneous catalyst" by Grasselli et al. [15]. According to this site isolation model, the presence of redox-inactive Ta as opposed to possibly redox-active, but unselective Mo at site 9 (S9) was proposed to play the role of isolating each hypothetical catalytic center comprised of one S2, two S4, two S7, and two S12 thereby improving the selectivity to acrylonitrile during propane ammoxidation.

The maximum selectivity to acrylonitrile observed during the propane ammoxidation reaction under a limited set of experimental conditions was plotted against the Ta S9 occupancy in MoVTeTaO M1 phase catalysts in Figure 6 of Supporting Information. The M1-Ta HT(0.12) showed the highest observed selectivity to acrylonitrile (~ 77 mol. %) among all MoVTeTaO M1 phase catalysts (Table 2). The highest selectivity to acrylonitrile for M1-Ta HT(0.12) is likely related to the highest Ta occupancy at S9 (~ 39 %) in this catalyst among all MoVTeTaO M1 phases investigated in this study, which is consistent with the site isolation model. However, there was no strong correlation between the Ta occupancy in S9 and the maximum selectivity toward acrylonitrile for all MoVTeTaO M1 phase catalysts investigated in this study. There is a possibility that we could not find the correlation between the Ta occupancy in S9 and the observed maximum selectivity toward acrylonitrile of MoVTeTaO M1 phase catalysts due to the lack of experimentally observed maximum selectivity to acrylonitrile during the reaction test. Another possible explanation is that there might be no relationship between the Ta occupancy in S9 and the experimentally observed highest selectivity to acrylonitrile of MoVTeTaO M1 phase catalysts in propane ammoxidation. This view may be supported by the recent study of Grasselli et al. [13], where $Mo_{0.6}V_{0.16-0.213}Nb_{0.055-0.10}Te_{0.10-0.18}O_X$ catalysts having different Nb/Te ratio showed the similar observed highest selectivity to acrylonitrile for the propane ammoxidation reaction (Figure 7 of Supporting Information). Further studies were conducted in order to establish the relationships between the Ta occupancy and observed maximum selectivity toward acrylonitrile for propane ammoxidation over MoVTeTaO M1 phase catalysts.

The Mo/V distribution at the proposed catalytic centers has been proposed to be one of the most important parameters governing the reactivity of the M1 phase in propane ammoxidation reaction [11]. First, we investigated the relationships between the Ta occupancy and Mo/V distribution in *ab* planes of M1 phase before further studying the relationships between the Mo/V distribution in *ab* planes of the M1 phase and its selectivity toward acrylonitrile in propane ammoxidation. The Mo/V ratio in M1 phase vs. Ta occupancy of S9 in all MoVTeTaO M1 phase catalysts is shown in Figure 8 of Supporting Information. We observed the general trend of increasing Ta occupancy in S9 along with increasing Mo/V ratios of M1 phase in Figure 8 of Supporting Information. These findings indicated that the V content of M1 phase decreased as the Ta occupancy in S9 increased. The possible scenario is explained as follows. S9 is fully occupied by Mo in the case of Nb-free MoVTeO M1 phase according to the findings of a previous study [11]. Ta is exclusively located in S9 in the case of the MoVTeTaO M1 phase catalysts as demonstrated in section 3.4.1. Ta employed during M1-Ta synthesis competes with Mo for S9, which forces Mo to other lattice sites, such as S1, 2, 3, 4, and 7, where it competes with V. As expected from such a scenario, increased Ta occupancy in S9 resulted in a decrease of V content of the M1 phase manifested in higher Mo/V ratios (Figure 8 of Supporting Information). Moreover, we further studied the correlation of Ta occupancy in site 9 with V content in the proposed catalytic center, on the basis of the probability model [13,14,25,42]. Therefore, the Mo/V ratios in the proposed catalytic center (S2-S4-S4-S7-S7) of the M1 phase were plotted against the Ta occupancy in S9 for all MoVTeTaO M1 phase catalysts in Figure 9 of Supporting Information. No strong correlation between Ta occupancy in S9 and the V contents in the catalytic center of M1 phase was observed in Figure 9 of Supporting Information. This result suggested that displaced Mo from S9 according to the scenario above compete with V located in the catalytic center, but do so without any preference for specific lattice sites in the catalytic center. This view is supported by the findings of Fu et al. [43] who simulated local configurations of MoVTeNbO M1 phase catalysts by a DFT-based method and evaluated the optimal Mo/V distribution in linking sites (S1, S2, S3, S4, and S7) on the basis of Boltzmann weighting factors. They proposed the following ranking of V occupancies: S2>S3>S7>S4≈S1 [43]. In summary, these findings suggested that Mo cations were displaced from S9 with increasing Ta content in the synthesis medium and competed with V cations for S1, S2, S3, S4, and S7. Next we examined the correlations between the acrylonitrile selectivity and Mo/V ratios on the basis of the probability model.

*3.5.2.2 Validation of probability model for MoVTeTaO M1 phase catalysts for propane ammoxidation*

The catalytic performance of MoVTeTaO M1 phase catalysts examined in this study was discussed above in section 3.5.1 (Table 2). In this section, we explore the correlations between the metal site occupancies in M1-Ta HT(0.09), M1-Ta HT(0.12), M1-Ta HT(0.15), M1-Ta MW, and M1-Ta HT(P) and their catalytic performance in propane ammoxidation. It has been suggested that the *ab* planes of M1 phase contain the active and selective sites for propane ammoxidation [42,44-46]. Our recent study of a MoVTeNbO M1 phase catalyst that selectively exposed *ab* planes demonstrated that the *ab* planes of M1 phase are responsible for high activity and selectivity in propane ammoxidation [47]. Furthermore, Grasselli et al. [13,14,25,42] developed a probability model where the Mo/V distribution in the proposed catalytic center present in the bulk *ab* planes of the M1 phase and consisting of sites 2, 4, 7, and 11 (S2-S4-S4-S7-S7), correlated with the selectivity to acrylonitrile. They advanced a hypothesis that one $V^{5+}$ cation present in the proposed catalytic center activates propane selectively, two $V^{5+}$ cations lead to waste products, while the absence of $V^{5+}$ cations in the proposed catalytic center corresponds to the lack of catalytic activity. Based on these assumptions, they estimated that the MoVTeNbO M1 phase catalyst contained 44% of active and selective sites for propane ammoxidation, 46% of inactive sites, and 10% of waste-forming sites. Using these estimates, they predicted the maximum acrylonitrile selectivity to be 81 mol. % (i.e., 44/(44+10) × 100%), which is somewhat higher than the maximum selectivity to acrylonitrile observed under a limited range of experimental conditions (72 mol. %) [42]. Therefore, this probability model suggested that the catalytic activity and selectivity for propane ammoxidation to acrylonitrile could be correlated with the Mo/V distribution at the proposed catalytic center.

In previous studies, the Mo/V distributions in the M1 phase have been characterized using a combination of NPD and S-XPD [32,48]. Due to the complexity of this mixed metal oxide, the structure refinement was very challenging and several versions of metal distributions in the MoVTeNbO M1 phase have been reported as discussed by Murayama et al. [48]. To overcome the limitations of powder diffraction methods, Pyrz et al. [9] employed the aberration corrected high resolution scanning transmission electron microscopy (STEM) and provided the atomic coordinates and lattice site occupancies based on direct observation of high angle annular dark-filed (HAADF) images of the MoVTeNb(Ta)O M1 phase. However, HAADF-STEM is also a bulk technique because each atomic column probed in the HAADF-STEM images contains hundreds of atoms, although the sample thickness in HAADF-STEM is considerably smaller as compared to other bulk techniques.

Our approach to determine the elemental composition of surface *ab* planes was introduced in order to address the limitations of powder diffraction methods [16,32,48,49] and further advance the HAADF-STEM methodology for the M1 catalytic system [8,9,11,50]. As discussed in section 3.4.2, the atomic column intensities of the bulk *ab* planes are very similar to those of surface *ab* planes based on the line profile HAADF-STEM imaging of M1-Ta HT (0.12) along the [hk0] directions. Given the observed similarities between the surface and bulk *ab* planes of M1-Ta HT (0.12), we proposed that the metal distributions in the bulk MoVTeTaO M1 phase determined by the HAADF-STEM imaging along the [001] direction also provide the compositional information relevant to that for the surface *ab* planes.

Accordingly, we applied the probability approach described by Pyrz et al. [11] to metal site distributions determined by our HAADF-STEM image analysis (Tables 1 and 2 of Supporting Information) in order to understand the relationships between the metal site distributions in MoVTeTaO M1 phase catalysts and their catalytic behavior in propane ammoxidation. Pyrz et al. [11] determined metal site occupancies in M1-Mo, M1-Nb, and M1-Ta variants (Table 4 of Supporting Information) and estimated the probabilities of finding 0-5 V cations in the proposed catalytic center (S2-S4-S4-S7-S7). According to their estimates, ~86% of MoVTeNbO, ~97% of MoVTeTaO, and 63% of MoVTeO M1 phases would have S2-S4-S4-S7-S7 centers containing less than 3 V cations (Table 5 of Supporting Information).

It is important to note that the probabilities reported by Pyrz et al. [11] and discussed in this section were calculated differently from those in the original probability model proposed by Grasselli et al. [13,14,25,42]. Pyrz et al. [11] did not account for the vanadium oxidation states, while the original Grasselli et al. probability model explicitly considered the probability of $V^{5+}$ in the proposed catalytic center (S2-S4-S4-S7-S7). Secondly, in the original probability model, S4 was fully occupied by Mo and, therefore, did not contribute to the $V^{5+}$ content in the catalytic center.

The Mo and V occupancies and the probabilities of finding 0-5 V cations in the catalytic center (S2-S4-S4-S7-S7) for all MoVTeTaO M1 phase catalysts are shown in Tables 6 and 7 of Supporting Information, respectively.

The probability, P, was calculated as shown below:

Probability of zero V cations in the catalytic center:

$P(V=0) = Mo2 \times Mo4 \times Mo4 \times Mo7 \times Mo7$, where Mo2, Mo4, and Mo7 are the partial occupancies in S2, S4, and S7, respectively (Table 6 of Supporting Information).

Probability of a single V cation in the catalytic center:

$P(V=1) = (V7 \times Mo7 \times Mo4 \times Mo4 \times Mo2) \times 2 + (V4 \times Mo4 \times Mo7 \times Mo7 \times Mo2) \times 2 + (V2 \times Mo4 \times Mo4 \times Mo7 \times Mo7)$, where V2, V4, and V7 are the partial occupancies in S2, S4, and S7, respectively (Table 6 of Supporting Information).

Probability of two V cations in the catalytic center:

$P(V=2) = (V7 \times V7 \times Mo4 \times Mo4 \times Mo2) + (V4 \times V4 \times Mo7 \times Mo7 \times Mo2) + (V4 \times V4 \times Mo7 \times M2) \times 4 + (V2 \times V4 \times Mo4 \times Mo7 \times Mo7) \times 2 + (V7 \times Mo2 \times Mo7 \times Mo4 \times Mo4) \times 2$.

Probability of three V cations in the catalytic center:

$P(V=3) = (V2 \times V4 \times V4 \times Mo7 \times Mo7) + (V2 \times Mo4 \times Mo4 \times V7 \times V7) + (V2 \times Mo4 \times V4 \times V7 \times Mo7) \times 4 + (Mo2 \times V4 \times V4 \times V7 \times Mo7) \times 2 + (Mo2 \times Mo4 \times V4 \times V7 \times V7) \times 2$

Probability of four V cations in the catalytic center:

$P(V=4) = (V2 \times V4 \times V4 \times V7 \times Mo7) \times 2 + (V2 \times Mo4 \times V4 \times V7 \times V7) \times 2 + (Mo2 \times V4 \times V4 \times V7 \times V7)$.

Probability of five V cations in the catalytic center:

$P(V=5) = (V2 \times V4 \times V4 \times V7 \times V7)$

According to the data shown in Table 8 of Supporting Information, no clear correlation was observed between the selectivity to acrylonitrile and the total V content in the proposed active center (S2-S4-S4-S7-S7). For example, M1-Ta HT(0.09) with the highest V content and M1-Ta HT(0.15) with the lowest V content in the active center showed the similar maximum selectivity to acrylonitrile. Moreover, no correlation was observed between the theoretical and experimental selectivities to acrylonitrile for the M1-Ta catalysts of this study (Table 8 of Supporting Information). Therefore, a new model is needed in order to establish the relationships between the catalytic center compositions and catalytic behavior of MoVTeTaO M1 phase catalysts.

*3.5.2.3 New probability models to predict propane ammoxidation behavior*

A recent study of Li et al. [49] indicated that in addition to sites 4 and 7 in the catalytic center, site 3 also contains $V^{5+}$, suggesting that propane can also be activated over site 3. Therefore, the first new model, Model 1, was proposed on the basis of the Grasselli et al. [13,14,25,42] probability model where S3 was included in the catalytic center. Moreover, the Li et al. study [49] also indicated that S2 contained only reduced $V^{4+}$, which was considered to be inactive in propane ammoxidation [42,54,55] and, therefore, S2 was excluded from our new probability model. Similar to previous models, we considered the new S3-S4-S4-S7-S7 center to be inactive when it contained no vanadium. The following assumptions were further made in order to calculate the probability of finding different numbers of $V^{5+}$ cations in this new S3-S4-S4-S7-S7 center: (1) the isolated $V^{5+}$ cation present in S3 only converts propane to propylene, (2) one $V^{5+}$ cation in the S4-S4-S7-S7 center converts propane (and the propylene intermediate) to ACN, (3) while two and more $V^{5+}$ cations in the S4-S4-S7-S7 center are propane combustion sites. Based on these assumptions, the probabilities of 0-1 $V^{5+}$ in S3 and 0-4 $V^{5+}$ in the new S4-S4-S7-S7 center were calculated (Table 9 of Supporting Information). According to our assumptions stated above, cases 1 and 5 shown in Table 9 of Supporting Information were considered to be only selective for propylene, while cases 2 and 3 were regarded to be selective for both propylene and acrylonitrile.

The sums of probabilities for cases 1 and 5, assumed to form only propylene, and for cases 2 and 3, assumed to form both propylene and ACN, for M1-Ta (0.09), M1-Ta (0.12), M1-Ta (0.15), M1-Ta MW, and M1-Ta HT(P) are shown in Table 10 of Supporting Information. The predicted maximum selectivities to acrylonitrile for all MoVTeTaO M1 phase catalysts, calculated by the method of Grasselli et al. [13,14,25,42], were also included in Table 10 of Supporting Information. According to Model 1, M1-Ta HT(0.12) and M1-Ta HT(0.15) are predicted to be more selective to acrylonitrile than M1-Ta HT(0.09) based on the fractions of selective sites shown in Table 10 of Supporting Information. However, there were no significant differences in the observed maximum selectivity to acrylonitrile between M1-Ta HT(0.15) and M1-Ta HT(0.09) in Table 2. Also, the predicted maximum selectivities to acrylonitrile for MoVTeTaO M1 phase catalysts shown in Table 10 of Supporting Information are considerably lower than the experimentally observed maximum selectivities reported in Table 2.

The combined yield of propylene and acrylonitrile was further plotted in Figure 10 of Supporting Information as a function of the probability of finding 1-2 $V^{5+}$ cations in the newly proposed S3-S4-S4-S7-S7 center since single $V^{5+}$ in S3 was proposed to only activate propane to propylene, while a single $V^{5+}$ in S4-S4-S7-S7 was proposed to not only convert propane to propylene, but also propylene to acrylonitrile. However, no correlation between the probability of finding 1-2 $V^{5+}$ in the newly proposed S3-S4-S4-S7-S7 center and combined yield of propylene and acrylonitrile was observed (Figure 10 of Supporting Information). Therefore, it appears that the Model 1 is unable to predict observed catalytic behavior of MoVTeTaO M1 phase catalysts in propane ammoxidation on the basis of metal distribution in *ab* planes of M1 phase.

Since site 3 was not found to be catalytically important based on the results of Model 1, Model 2 was considered next where this site was excluded. Moreover, Model 2 attempted to correlate the catalytic performance with the total V content in the catalytic center (S2-S4-S4-S7-S7), irrespective of its oxidation state. The yield of acrylonitrile was plotted in Figure 3 as a function of total V content in the catalytic center (S2-S4-S4-S7-S7). Recently, Naraschewski et al. [51,52] reported a linear relationship between the initial rate of propane conversion and V content of MoVTeNbO catalysts observed in propane oxidation to acrylic acid. A similar correlation was found in this study for the yield of acrylonitrile during propane ammoxidation over MoVTeTaO M1 phase catalysts which increased with the total V content in S2-S4-S4-S7-S7 (Figure 3) with one exception of M1-Ta HT(P). Possible reasons for this catalyst being an outlier are discussed further below.

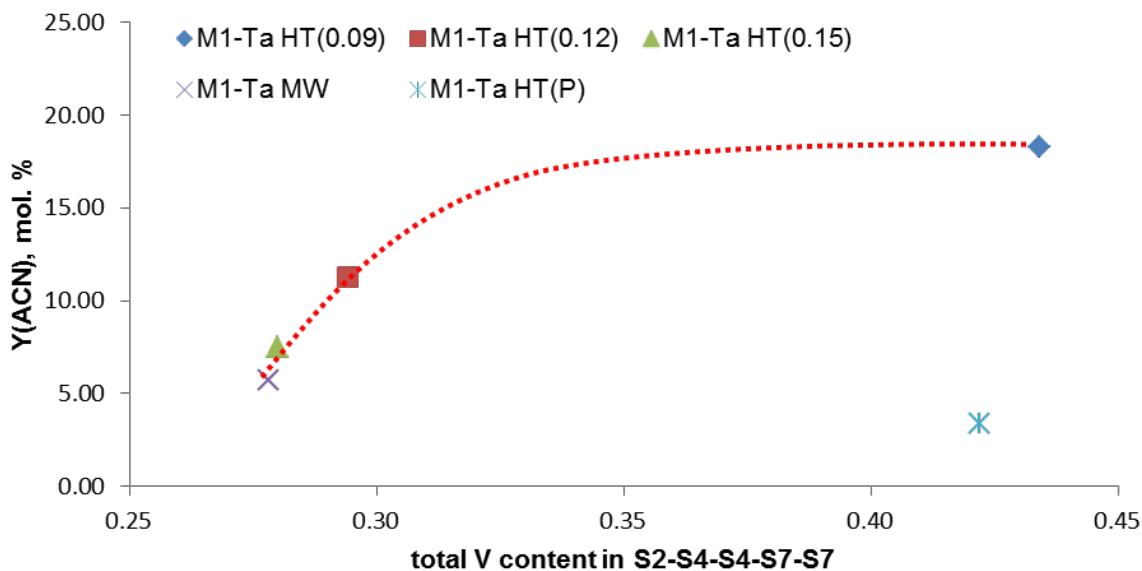

**Figure 3.** Yield of acrylonitrile, Y(ACN), as a function of total V content in the catalytic center (S2-S4-S4-S7-S7) of M1-Ta HT(0.09), M1-Ta HT(0.12), M1-Ta HT(0.15), M1-Ta MW, and HT(P) in propane ammoxidation; Reaction conditions: $C_3H_8$:$NH_3$:$O_2$:He=5.7:8.6:17.1:68.6 (%); total flow rate, 26.3 mL•$min^{-1}$; 0.2 g catalyst; reaction temperature: 420°C.

We further attempted to correlate the catalytic activity in propane consumption with the Mo/V distribution in *ab* planes of M1 phase on the basis of the probability Model 2. According to previous kinetic studies of MoVTeNbO M1 catalyst in propane ammoxidation, the rate of propane consumption over the MoVTeTaO M1 phase catalysts under propane ammoxidation conditions was fit well by a *first-order irreversible* reaction kinetics for a plug flow reactor model [4,56,57]. The reaction data for propane ammoxidation over MoVTeTaO M1 phase catalysts are summarized in Table 2. The Arrhenius plots of the first-order irreversible reaction rate constant, k", for MoVTeTaO M1 phase catalysts were linear supporting the conclusion that the rate of propane consumption is the 1st order reaction (Figure 11 of Supporting Information). The activation energies of propane consumption over MoVTeTaO M1 phase catalysts from the Arrhenius plots (Figure 11 of Supporting Information) are in the range of 112~158 (kJ·$mol^{-1}$).

These activation energies are in good agreement with the activation energy (131 kJ·mol$^{-1}$) of MoVTeNbO M1 phase reported in the previous study where propane ammoxidation was also assumed to be a first-order reaction [4].

The propane consumption rate constants, k", for all catalysts are plotted against the total V content in the catalytic center (S2-S4-S4-S7-S7) of M1 phase in Figure 12 of Supporting Information, which showed a lack of correlation. It may be explained by the fact that k" values have been normalized to the total BET surface area of the M1 catalysts, whereas the catalytic activity in propane ammoxidation is thought to be associated with the presence of only surface *ab* planes. This view is supported by the results of our recent study of a MoVTeNbO M1 phase selectively exposing *ab* planes which provided additional evidence that the *ab* planes may be responsible for its high activity and selectivity in propane ammoxidation [47]. The MoVTeTaO M1 phase catalysts employed in this study showed rod-shaped crystal morphologies with different extent of exposure of surface *ab* planes (Figure 2 of Supporting Information). In order to account for these differences, the reaction rate constants, k", were normalized to the surface *ab* plane areas of these M1 phases estimated by measuring the lengths and diameters of statistically representative numbers (N=30) of M1 phase crystals in respective SEM images of each catalyst (Table 11 of Supporting Information). The *ab* plane surface areas were estimated based on three assumptions: all MoVTeTaO catalysts in this study were (1) pure M1 phase, (2) 100% crystalline, (3) and possessed a perfect cylindrical shape.

It is worth mentioning that the length of M1 crystals decreased with the Ta synthesis concentration for M1-Ta HT(0.09), M1-Ta HT(0.12), and M1-Ta HT(0.15). This observation is in good agreement with previous findings that Nb$^{5+}$ (and oxalate) ions suppressed crystal growth of the MoVTeNbO M1 phase [58]. The estimated surface *ab* plane areas were used to calculate the new reaction rate constant, k"$_{ab}$. Those k"$_{ab}$ were plotted against the total V content in the catalytic center (S2-S4-S4-S7-S7) of MoVTeTaO M1 phase catalysts in Figure 4.

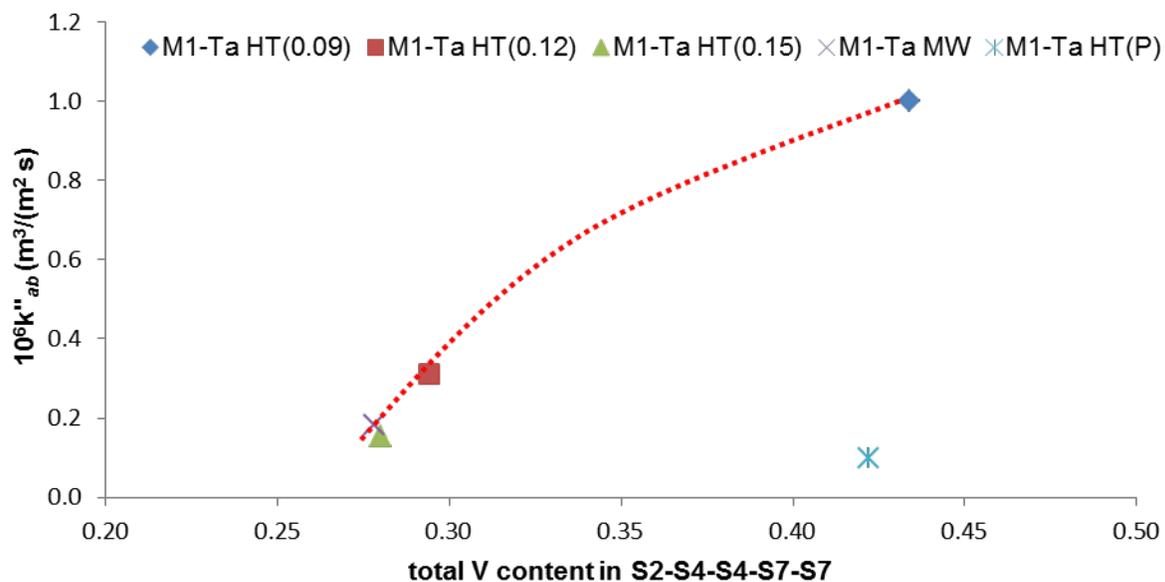

**Figure 4.** Reaction rate constant, k"$_{ab}$, normalized to the estimated *ab* plane surface areas vs. the total V content in the catalytic center (S2-S4-S4-S7-S7) of M1-Ta HT(0.09), M1-Ta HT(0.12), M1-Ta HT(0.15), M1-Ta MW, and HT(P) in propane ammoxidation; Reaction conditions: $C_3H_8$:$NH_3$:$O_2$:He=5.7:8.6:17.1:68.6; total flow rate, 26.3 mL•min$^{-1}$; 0.2 g catalyst; reaction temperature: 420°C.

The reaction rate constant, k"$_{ab}$, was found to increase with the total V content in the catalytic center (S2-S4-S4-S7-S7) of M1 phase catalysts (Figure 4), with one exception of M1-Ta HT(P), which is discussed further below. These observations are consistent with the results of recent studies that showed a linear relationship between the initial rate of propane conversion and the vanadium content of MoVTeNbO M1 phase in propane oxidation to acrylic acid [51,52].

In summary, the new probability model 2 was relatively successful in establishing a correlation between the yield of acrylonitrile and the total V content in the proposed catalytic center of M1 phase (Figure 3) and a correlation between the reaction rate constant, k"$_{ab}$, and the total V content in the proposed catalytic center of M1 phase (Figure 4). These findings are also consistent with the conclusions of the probability model proposed by Grasselli et al. [13,14,25,42] and other studies suggesting the importance of V content in the M1 phase for propane activation during propane ammoxidation [51,52,59].

Although the new probability Model 2 established a correlation between the catalytic behavior and the total V content in the catalytic center of the MoVTeTaO M1 phase catalysts of this study, it did not probe the importance of multiple vanadium sites for propane ammoxidation, which were previously suggested to be more efficient in propane activation [60-62]. Indeed, a previous experimental study of propane oxidation to acrylic acid combined with chemical probe chemisorption and low energy ion scattering (LEIS) conducted by our group indicated that multiple $VO_x$ sites present in M1 phases may be more efficient in selective propane oxidation than isolated $VO_x$ species [60]. Therefore, another probability model, Model 3, was explored in order to improve the predictive capability for all MoVTeTaO M1 phase catalysts employed in this study. Since Model 2 established that the reactivity in propane ammoxidation correlated with the total V content in the catalytic center, Model 3 was proposed to test the hypothesis that the activity and selectivity in propane ammoxidation were correlated with presence of more than one V in the catalytic center (S2-S4-S4-S7-S7) of M1 phase.

In Model 3, the catalytic reactivity in propane ammoxidation was correlated with: (1) P(V=2), the probability of two V in the catalytic center (case 1); (2) P(V=2)+P(V=3), the combined probability of two V and three V in the catalytic center (case 2); and (3) P(V=2)+P(V=3)+P(V=4), the combined probability of two V, three V, and four V in the catalytic center (case 3). Although the probability of five V in the catalytic center is theoretically possible, the P(V=5) was significantly below 0.1 for all MoVTeTaO M1 phase catalysts and, therefore, was not included in this study.

The probabilities of P(V>2) in the catalytic center (S2-S4-S4-S7-S7) shown in Table 3 were calculated according to equations shown in section 3.5.2.2. The probabilities of V for case 3 for all MoVTeTaO M1 phases were plotted against the yield of acrylonitrile in Figure 5. As shown in Figure 5, the yield of acrylonitrile increased with the probability of V>2 in the catalytic center except M1-Ta HT(P), which is discussed further below.

**Table 3.** Probabilities of the number of V cations more than 2, P(V>2), in the catalytic center (S2-S4-S4-S7-S7) of M1-Ta (0.09), M1-Ta (0.12), M1-Ta (0.15), M1-Ta MW, and M1-Ta HT(P).

|  | M1-Ta HT (0.09) | M1-Ta HT(0.12) | M1-Ta HT (0.15) | M1-Ta MW | M1-Ta HT(P) |
|---|---|---|---|---|---|
| Case 1: P(V=2) | 0.36 | 0.33 | 0.33 | 0.31 | 0.36 |
| Case 2: P(V=2)+P(V=3) | 0.62 | 0.44 | 0.43 | 0.41 | 0.61 |
| Case 3: P(V=2)+P(V=3)+P(V=4) | 0.72 | 0.46 | 0.44 | 0.43 | 0.70 |

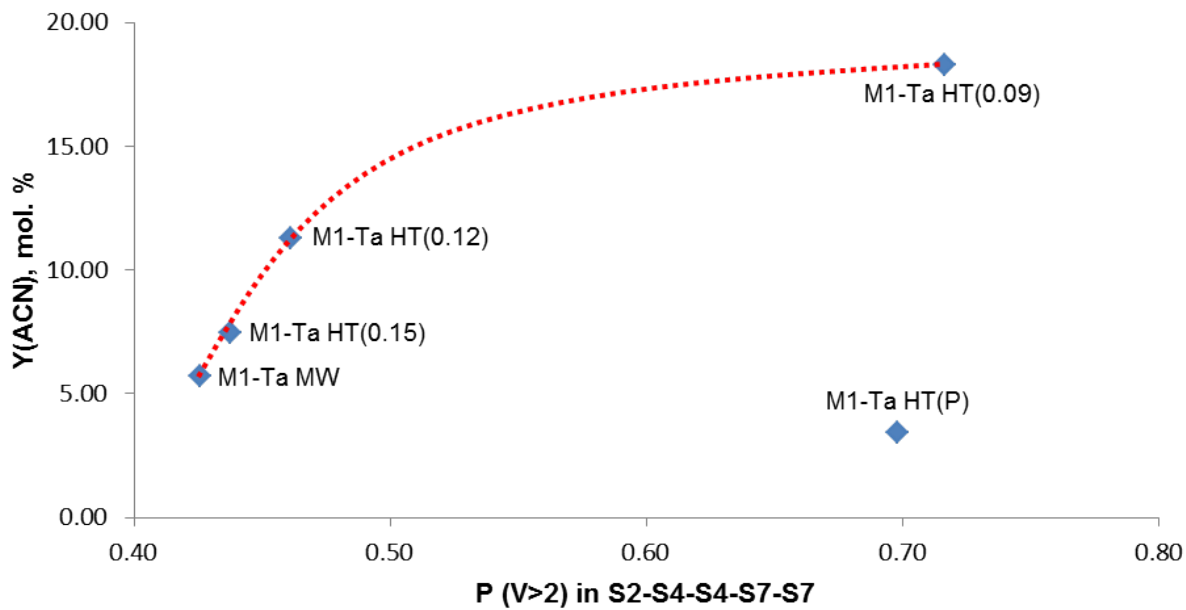

**Figure 5.** Yield of acrylonitrile, Y(ACN), as a function of probability of V>2 in the catalytic center (S2-S4-S4-S7-S7) of M1-Ta HT(0.09), M1-Ta HT(0.12), M1-Ta HT(0.15), M1-Ta MW, and HT(P) in propane ammoxidation; Reaction conditions: $C_3H_8$:$NH_3$:$O_2$:He=5.7:8.6:17.1:68.6; total flow rate, 26.3 mL•$min^{-1}$; 0.2 g catalyst; reaction temperature: 420°C.

The propane consumption rate constants, k", based on BET surface areas are plotted against the probabilities of V>2 in the catalytic center (S2-S4-S4-S7-S7) for all MoVTeTaO M1 phase catalysts in Figure 13 of Supporting Information. Similar to Model 2, no correlations between the propane consumption rate constants, k", and probabilities of V>2 in the catalytic center are observed. Therefore, the reaction rate constants, $k"_{ab}$, based on the geometric estimates of surface *ab* plane areas (Table 11 of Supporting Information) were plotted against the probabilities of V>2 in the catalytic center (Figure 6), which showed similar increasing trends of $k"_{ab}$ with the probabilities of V>2 in the catalytic center to those for Model 2 shown in Figure 4. The observed correlations between $k"_{ab}$ and the probabilities of V>2 in S2-S4-S4-S7-S7 lend further support to our previous findings [60-62] that multiple $VO_x$ redox sites present in the surface *ab* planes of the M1 phase may be more active in propane (amm)oxidation.

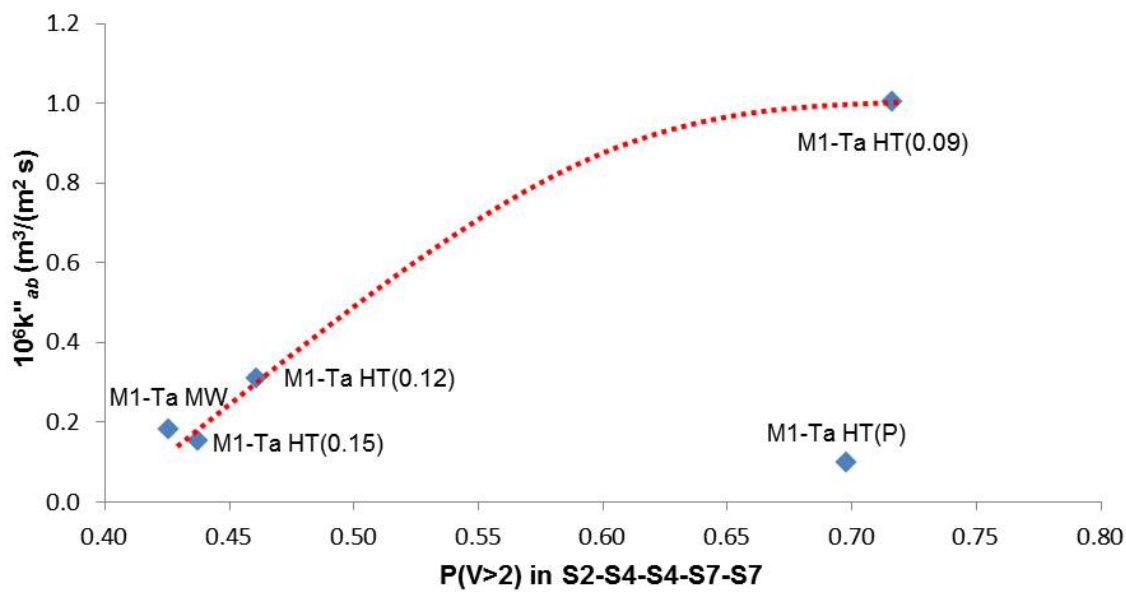

**Figure 6.** Reaction rate constant, $k''_{ab}$, based on the estimated *ab* planes surface area plotted as a function of probability of V>2 in the catalytic center (S2-S4-S4-S7-S7) of M1-Ta HT(0.09), M1-Ta HT(0.12), M1-Ta HT(0.15), M1-Ta MW, and HT(P) in propane ammoxidation; Reaction conditions: $C_3H_8:NH_3:O_2:He=5.7:8.6:17.1:68.6$; total flow rate, 26.3 mL•min$^{-1}$; 0.2 g catalyst; reaction temperature: 420°C.

However, the new probability models were unable to explain the catalytic behavior of M1-Ta HT(P), which was an outlier according to Model 2 and Model 3. These observations may be understood in terms of the detrimental impact of oxalate anion on the oxidation state of Te in the synthesis gel. In a recent study, Ivars et al. [58,63] prepared the MoVTeNbO M1 phases at different oxalate/Mo molar ratios in the synthesis gel and found that the most selective catalysts for propane oxidation possessed the oxalate/Mo ratios of 0.4~0.6. They proposed that the oxalate anion was responsible for $Te^{6+}$ reduction to $Te^{4+}$ and suggested that controlling the oxalate concentration in the synthesis gel is critical to obtaining highly active and selective M1 phases for propane oxidation. Therefore, in light of these earlier findings [58,63], we speculate that the low reactivity of M1-Ta HT(P) in propane ammoxidation may be explained by excessive $Te^{6+}$ reduction to $Te^{4+}$ due to the presence of oxalate during synthesis.

## 4. Conclusions

In this study, the MoVTeTaO M1 phase catalysts were successfully synthesized by conventional hydrothermal (HT) and microwave-assisted (MW) approaches employing Ta ethoxide and Ta oxalate as Ta sources. The XRD patterns of all catalysts after the $H_2O_2$ treatment showed the presence of pure M1 phase. It was found that all catalysts were active in propane ammoxidation and showed different selectivity to acrylonitrile depending on synthesis methods and compositions. The HAADF-STEM image analysis of all MoVTeTaO M1 phases provided metal distributions in various lattice sites and confirmed that Ta is exclusively located in S9. The profile intensity analysis of the M1 phase oriented along [hk0] directions from the surface to bulk region indicated that the chemical composition of surface *ab* planes is very similar to their composition in the bulk. The HAADF-STEM image analysis showed that synthesis methods have a significant impact on the Mo/V distribution as well as Ta occupancy of MoVTeTaO M1 phase catalysts examined in this study. M1-Ta HT prepared by a hydrothermal synthesis method exhibited the highest Ta occupancy at S9. A general trend of decreasing of V content with increasing Ta occupancy in S9 was observed. The obtained Mo/V distributions were initially employed to predict theoretical selectivities to acrylonitrile according to earlier probability models, but no clear correlations were obtained with observed catalytic behavior of the MoVTeTaO M1 phases in propane ammoxidation.

Three new probability models, Models 1, 2, and 3, were tested in this study. Model 1, where S3 was incorporated into the S3-S4-S4-S7-S7 center, was unable to correlate the catalytic performance of MoVTeTaO M1 phase catalysts in propane ammoxidation with their Mo/V distributions. However, Models 2 and 3 suggested correlations between new descriptors of the Mo/V distribution in the *ab* planes of M1 phase and its catalytic behavior. Specifically, enhanced acrylonitrile yield and 1$^{st}$ order irreversible reaction rate constants for propane consumption, k"$_{ab}$, normalized to the estimated surface *ab* plane areas, correlated with increasing the total V content in the S2-S4-S4-S7-S7 (Model 2) or the probability of V over 2 in the S2-S4-S4-S7-S7 (Model 3). The correlations for k" were observed only when k" was normalized to the surface *ab* plane areas (as opposed to the total BET surface areas), lending further support to the idea that the surface *ab* planes may contain the active and selective surface sites for propane ammoxidation. Model 3 resulted in a correlation emphasizing the importance of multiple VO$_x$ sites for the catalytic activity and selectivity in propane ammoxidation. Moreover, the improved understanding of fundamental relationships between the composition of *ab* planes of MoVTeTaO M1 phase and its catalytic behavior can provide rules of rational design of improved mixed metal oxide catalysts for propane ammoxidation and other selective oxidation reactions.

**Acknowledgements**

This study was supported by the Chemical Sciences, Geosciences and Biosciences Division, Office of Basic Energy Sciences, U.S. Department of Energy, under Grant #DE-FG02-04ER15604. Electron microscopy research is supported by the U.S. Department of Energy, Office of Science, Basic Energy Sciences, Materials Sciences and Engineering Division and through a user project supported by ORNL's Center for Nanophase Materials Sciences, sponsored by the Scientific User Facilities Division, Office of Science, Basic Energy Sciences,

U.S. Department of Energy. The authors would like to thank Shawn Reeves (ORNL) for her help with preparing sectioned samples. The resources of the National Energy Research Scientific Computing Center, supported by DOE Office of Science under Contract DE-AC02-05CH11231, are gratefully acknowledged.

# Supporting Information

**Legends for Tables**

Table S1. Metal site occupancies of the M1-Ta HT(0.12), MW, and HT(P) catalysts calculated from HAADF-STEM images.

Table S2. Metal occupancies of the M1-Ta HT(0.09), HT(0.12), and HT(0.15) catalysts calculated from HAADF-STEM images.

Table S3. Comparison of metal ratios for M1-Ta HT(0.09), M1-Ta HT(0.12) and M1-Ta HT(0.15) catalysts.

Table S4. STEM-based Mo and V occupancies in the proposed catalytic center (S2-S4-S4-S7-S7) of MoVTeO, MoVTeNbO, and MoVTeTaO M1 phase catalysts [1].

Table S5. The calculated probability that the proposed catalytic center (S2-S4-S4-S7-S7) in the MoVTeO, MoVTeNbO, and MoVTeTaO M1 phase catalysts will contain between zero and five vanadium cations [1].

Table S6. STEM-based Mo and V occupancies in the proposed catalytic center (S2-S4-S4-S7-S7) of M1-Ta HT(0.09), M1-Ta HT(0.12), M1-Ta HT(0.15), M1-Ta MW, and M1-Ta HT(P).

Table S7. The probabilities of finding 0-5 V cations in the proposed catalytic center (S2-S4-S4-S7-S7) in M1-Ta HT(0.09), M1-Ta HT(0.12), M1-Ta HT(0.15), M1-Ta MW, and M1-Ta HT(P).

Table S8. Fractions of active centers and V content in the catalytic center (S2-S4-S4-S7-S7) of M1-Ta HT(0.09), M1-Ta HT(0.12), M1-Ta HT(0.15), M1-Ta MW, and M1-Ta HT(P) along with the experimentally observed maximum selectivity to acrylonitrile.

Table S9. Probabilities of finding 0-4 $V^{5+}$ cations in the newly proposed catalytic center (S3-S4-S7) for M1-Ta (0.09), M1-Ta (0.12), M1-Ta (0.15), M1-Ta MW, and M1-Ta HT(P).

Table S10. The sums of probabilities for cases that are assumed to form only propylene and ACN for M1-Ta (0.09), M1-Ta (0.12), M1-Ta (0.15), M1-Ta MW, and M1-Ta HT(P).

Table S11. Geometric estimates of surface *ab* plane areas based on the analysis of M1 phase crystals in SEM images.

**Table S1.** Metal site occupancies of the M1-Ta HT(0.12), MW, and HT(P) catalysts calculated from HAADF-STEM images.

| | M1-Ta HT(0.12) | | | | M1-Ta MW | | | | M1-Ta HT(P) | | | | M1-Ta SE [2] | | |
|---|---|---|---|---|---|---|---|---|---|---|---|---|---|---|---|
| | Mo | V | Ta | C.I* | Mo | V | Ta | C.I* | Mo | V | Ta | C.I* | Mo | V | Ta |
| S1 | 0.84 | 0.16 | | 0.12 | 1 | 0 | | 0.09 | 0.56 | 0.44 | | 0.08 | 1.00 | 0.00 | |
| S2 | 0.39 | 0.61 | | 0.05 | 0.41 | 0.59 | | 0.04 | 0.35 | 0.65 | | 0.05 | 0.37 | 0.63 | |
| S3 | 0.63 | 0.37 | | 0.04 | 0.58 | 0.42 | | 0.03 | 0.53 | 0.47 | | 0.05 | 0.34 | 0.66 | |
| S4 | 0.87 | 0.13 | | 0.09 | 0.89 | 0.11 | | 0.07 | 0.7 | 0.3 | | 0.07 | 1.00 | 0.00 | |
| S5 | 0.99 | 0.01 | | 0.01 | 1.02 | (0.02) | | 0.01 | 1.01 | (0.01) | | 0.02 | 1.00 | 0.00 | |
| S6 | 0.95 | 0.05 | | 0.05 | 0.9 | 0.1 | | 0.03 | 0.95 | 0.05 | | 0.03 | 1.00 | 0.00 | |
| S7 | 0.7 | 0.3 | | 0.11 | 0.71 | 0.29 | | 0.06 | 0.57 | 0.43 | | 0.06 | 0.59 | 0.41 | |
| S8 | 1.01 | (0.01) | | 0.01 | 0.98 | 0.02 | | 0.01 | 0.99 | 0.01 | | 0.02 | 1.00 | 0.00 | |
| S9 | 0.61 | | 0.39 | 0.04 | 0.67 | | 0.33 | 0.03 | 0.79 | | 0.21 | 0.04 | 0.56 | | 0.44 |
| S10 | 0.99 | 0.01 | | 0.04 | 1.01 | (0.01) | | 0.03 | 0.93 | 0.07 | | 0.03 | 1.00 | 0.00 | |
| S11 | 1.02 | (0.02) | | 0.05 | 1.14 | (0.14) | | 0.06 | 0.95 | 0.05 | | 0.04 | 1.00 | 0.00 | |
| Sum | 9 | 1.6 | | | 9.29 | 1.38 | | | 8.33 | 2.46 | | | 8.85 | 1.71 | |

C.I* is upper and lower 95% confidence limits; ( ) indicates a negative value.

**Table S2**. Metal occupancies of the M1-Ta HT(0.09), HT(0.12), and HT(0.15) catalysts calculated from HAADF-STEM images.

| | M1-Ta HT(0.09) | | | | M1- Ta HT(0.12) | | | | M1- Ta HT(0.15) | | | |
|---|---|---|---|---|---|---|---|---|---|---|---|---|
| | Mo | V | Ta | C.I* | Mo | V | Ta | C.I* | Mo | V | Ta | C.I* |
| S1 | 0.55 | 0.45 | | 0.15 | 0.84 | 0.16 | | 0.12 | 0.63 | 0.37 | | 0.16 |
| S2 | 0.33 | 0.67 | | 0.12 | 0.39 | 0.61 | | 0.05 | 0.42 | 0.58 | | 0.15 |
| S3 | 0.56 | 0.44 | | 0.11 | 0.63 | 0.37 | | 0.04 | 0.47 | 0.53 | | 0.09 |
| S4 | 0.69 | 0.31 | | 0.21 | 0.87 | 0.13 | | 0.09 | 0.94 | 0.06 | | 0.31 |
| S5 | 1.02 | (0.02) | | 0.07 | 0.99 | 0.01 | | 0.01 | 1.04 | (0.04) | | 0.06 |
| S6 | 0.94 | 0.06 | | 0.11 | 0.95 | 0.05 | | 0.05 | 1 | 0 | | 0.14 |
| S7 | 0.56 | 0.44 | | 0.1 | 0.7 | 0.3 | | 0.11 | 0.65 | 0.35 | | 0.12 |
| S8 | 0.98 | 0.02 | | 0.07 | 1.01 | (0.01) | | 0.01 | 0.96 | 0.04 | | 0.06 |
| S9 | 0.88 | | 0.12 | 0 | 0.61 | | 0.39 | 0.04 | 0.8 | | 0.2 | 0.07 |
| S10 | 1.02 | (0.02) | | 0.11 | 0.99 | 0.01 | | 0.04 | 0.82 | 0.18 | | 0.15 |
| S11 | 0.99 | 0.01 | | 0.14 | 1.02 | (0.02) | | 0.05 | 0.99 | 0.01 | | 0.11 |
| Sum | 8.52 | 2.36 | | | 9 | 1.61 | | | 8.72 | 2.08 | | |

C.I* is upper and lower 95% confidence limits; ( ) indicates a negative value.

**Table S3**. Comparison of metal ratios for M1-Ta HT(0.09), M1-Ta HT(0.12) and M1-Ta HT(0.15) catalysts.

|  | M1-Ta HT(0.09) | | M1-Ta HT(0.12) | | M1-Ta HT(0.15) | |
|---|---|---|---|---|---|---|
|  | Mo/V | Mo/Ta | Mo/V | Mo/Ta | Mo/V | Mo/Ta |
| Synthesis | 3.2 | 11.1 | 3.2 | 8.3 | 3.2 | 6.7 |
| SEM/EDS (Bulk) | 3.6 | 3.6 | 3.1 | 3.0 | 3.3 | 2.9 |
| HAADF-STEM imaging | 3.6 | 71.0 | 5.6 | 23.1 | 4.2 | 43.6 |

**Table S4.** STEM-based Mo and V occupancies in the proposed catalytic center (S2-S4-S4-S7-S7) of MoVTeO, MoVTeNbO, and MoVTeTaO M1 phase catalysts [1].

|  | MoVTeO | | MoVTeNbO | | MoVTeTaO | |
|---|---|---|---|---|---|---|
|  | Mo | V | Mo | V | Mo | V |
| S2 | 0.57 | 0.43 | 0.5 | 0.5 | 0.42 | 0.58 |
| S4 | 0.62 | 0.38 | 0.79 | 0.21 | 0.95 | 0.05 |
| S7 | 0.53 | 0.47 | 0.74 | 0.26 | 0.85 | 0.15 |
| Total | 1.72 | 1.28 | 2.03 | 0.97 | 2.22 | 0.78 |

**Table S5.** The calculated probability that the proposed catalytic center (S2-S4-S4-S7-S7) in the MoVTeO, MoVTeNbO, and MoVTeTaO M1 phase catalysts will contain between zero and five vanadium cations [1].

| # V cations | MoVTeO | MoVTeNbO | MoVTeTaO |
|---|---|---|---|
| 0 | 0.06 | 0.17 | 0.27 |
| 1 | 0.23 | 0.38 | 0.50 |
| 2 | 0.34 | 0.31 | 0.19 |
| 3 | 0.26 | 0.12 | 0.03 |
| 4 | 0.09 | 0.02 | 0 |
| 5 | 0.01 | 0 | 0 |
| S2-S4-S4-S7-S7 centers containing 0-2 V | 0.63 | 0.86 | 0.96 |

**Table S6.** STEM-based Mo and V occupancies in the proposed catalytic center (S2-S4-S4-S7-S7) of M1-Ta HT(0.09), M1-Ta HT(0.12), M1-Ta HT(0.15), M1-Ta MW, and M1-Ta HT(P).

| | M1-Ta HT (0.09) | | M1-Ta HT (0.12) | | M1-Ta HT (0.15) | | M1-Ta MW | | M1-Ta HT(P) | |
|---|---|---|---|---|---|---|---|---|---|---|
| | Mo | V | Mo | V | Mo | V | Mo | V | Mo | V |
| S2 | 0.33 | 0.67 | 0.39 | 0.61 | 0.42 | 0.58 | 0.41 | 0.59 | 0.35 | 0.65 |
| S4 | 0.69 | 0.31 | 0.87 | 0.13 | 0.94 | 0.06 | 0.89 | 0.11 | 0.7 | 0.3 |
| S7 | 0.56 | 0.44 | 0.7 | 0.3 | 0.65 | 0.35 | 0.71 | 0.29 | 0.57 | 0.43 |
| Total | 1.58 | 1.42 | 1.96 | 1.04 | 2.01 | 0.99 | 2.01 | 0.99 | 1.62 | 1.38 |

**Table S7.** The probabilities of finding 0-5 V cations in the proposed catalytic center (S2-S4-S4-S7-S7) in M1-Ta HT(0.09), M1-Ta HT(0.12), M1-Ta HT(0.15), M1-Ta MW, and M1-Ta HT(P).

| # of V atoms in the active center | M1-Ta HT (0.09) | M1-Ta HT (0.12) | M1-Ta HT (0.15) | M1-Ta MW | M1-Ta HT(P) |
|---|---|---|---|---|---|
| 0 | 0.05 | 0.14 | 0.16 | 0.16 | 0.06 |
| 1 | 0.22 | 0.39 | 0.41 | 0.41 | 0.24 |
| 2 | 0.36 | 0.33 | 0.33 | 0.31 | 0.36 |
| 3 | 0.27 | 0.12 | 0.10 | 0.10 | 0.25 |
| 4 | 0.09 | 0.02 | 0.01 | 0.01 | 0.08 |
| 5 | 0.01 | 0.00 | 0.00 | 0.00 | 0.01 |
| S2-S4-S4-S7-S7 centers containing 0-2 V | 0.63 | 0.86 | 0.90 | 0.88 | 0.66 |

**Table S8.** Theoretical selectivity to acrylonitrile and V content in the S2-S4-S4-S7-S7 catalytic center of M1-Ta HT(0.09), M1-Ta HT(0.12), M1-Ta HT(0.15), M1-Ta MW, and M1-Ta HT(P) along with the experimental maximum selectivity to acrylonitrile.

| | *Theoretical selectivity (%) | V content in the active center | ‡Maximum selectivity to acrylonitrile (%) |
|---|---|---|---|
| M1-Ta HT(0.09) | 23 | 1.42 | 71.6 |
| M1-Ta HT(0.12) | 45 | 1.04 | 76.9 |
| M1-Ta HT(0.15) | 49 | 0.99 | 70.5 |
| M1-Ta MW | 49 | 0.99 | 70.4 |
| M1-Ta HT(P) | 26 | 1.38 | 68.9 |

*Theoretical selectivity to acrylonitrile is defined as 100% x P(1V)/(1-P(0V)), where P(1V) and P(0V) are probabilities of finding a single and zero V cations, respectively, in S2-S4-S4-S7-S7; ‡ experimental maximum selectivity to acrylonitrile.

**Table S9.** Probabilities of finding 0-4 $V^{5+}$ cations in the newly proposed catalytic center (S3-S4-S4-S7-S7) for M1-Ta (0.09), M1-Ta (0.12), M1-Ta (0.15), M1-Ta MW, and M1-Ta HT(P).

| Case 1) | Catalyst | $V^{5+}$=1 in S3 | $V^{5+}$=0 in the "active center" |
|---|---|---|---|
| | M1-Ta HT(0.09) | 0.44 | 0.15 |
| | M1-Ta HT(0.12) | 0.37 | 0.37 |
| | M1-Ta HT(0.15) | 0.53 | 0.37 |
| | M1-Ta MW | 0.42 | 0.40 |
| | M1-Ta HT(P) | 0.47 | 0.16 |
| Case 2) | Catalyst | $V^{5+}$=0 in S3 | $V^{5+}$=1 in the "active center" |
| | M1-Ta HT(0.09) | 0.56 | 0.37 |
| | M1-Ta HT(0.12) | 0.63 | 0.43 |
| | M1-Ta HT(0.15) | 0.47 | 0.45 |
| | M1-Ta MW | 0.58 | 0.42 |
| | M1-Ta HT(P) | 0.53 | 0.38 |
| Case 3) | Catalyst | $V^{5+}$=1 in S3 | $V^{5+}$=1 in the "active center" |
| | M1-Ta HT(0.09) | 0.44 | 0.37 |
| | M1-Ta HT(0.12) | 0.37 | 0.43 |
| | M1-Ta HT(0.15) | 0.53 | 0.45 |
| | M1-Ta MW | 0.42 | 0.42 |
| | M1-Ta HT(P) | 0.47 | 0.38 |
| Case 4) | Catalyst | $V^{5+}$=0 in S3 | $V^{5+}$>1 in the "active center" |
| | M1-Ta HT(0.09) | 0.56 | 0.48 |
| | M1-Ta HT(0.12) | 0.63 | 0.20 |
| | M1-Ta HT(0.15) | 0.47 | 0.18 |
| | M1-Ta MW | 0.58 | 0.18 |
| | M1-Ta HT(P) | 0.53 | 0.46 |
| Case 5) | Catalyst | $V^{5+}$=1 in S3 | $V^{5+}$>1 in the "active center" |
| | M1-Ta HT(0.09) | 0.44 | 0.48 |
| | M1-Ta HT(0.12) | 0.37 | 0.20 |
| | M1-Ta HT(0.15) | 0.53 | 0.18 |
| | M1-Ta MW | 0.42 | 0.18 |
| | M1-Ta HT(P) | 0.47 | 0.46 |
| Case 6) | Catalyst | $V^{5+}$=0 in S3 | $V^{5+}$=0 in the "active center" |
| | M1-Ta HT(0.09) | 0.56 | 0.15 |
| | M1-Ta HT(0.12) | 0.63 | 0.37 |
| | M1-Ta HT(0.15) | 0.47 | 0.37 |
| | M1-Ta MW | 0.58 | 0.40 |
| | M1-Ta HT(P) | 0.53 | 0.16 |

**Table S10**. The sums of probabilities for cases that are assumed to form only propylene (cases 1 and 5) and ACN (cases 2 and 3) for M1-Ta (0.09), M1-Ta (0.12), M1-Ta (0.15), M1-Ta MW, and M1-Ta HT(P).

|  | selective (propylene) | selective (ACN) | *Predicted maximum selectivity to ACN, % |
|---|---|---|---|
| M1-Ta HT(0.09) | 0.28 | 0.37 | 40 |
| M1-Ta HT(0.12) | 0.21 | 0.43 | 56 |
| M1-Ta HT(0.15) | 0.29 | 0.45 | 55 |
| M1-Ta MW | 0.24 | 0.42 | 55 |
| M1-Ta HT(P) | 0.29 | 0.38 | 41 |

* calculated as 100% ×Σ selective centers to ACN / Σ [selective centers to ACN + propylene + waste centers (case 4 in Table S9)]. It is assumed that all propylene, formed at the selective centers to ACN, is converted to ACN.

**Table S11**. Geometric estimates of surface *ab* plane areas based on the analysis of M1 phase crystals in SEM images.

|  | $l_{mean}$ (nm) | $d_{mean}$ (nm) | R (nm) | C.*ab*.S (m$^2$/g) |
|---|---|---|---|---|
| MoVTeTaO HT(0.09) | 265±24 | 31±4 | 16 | 1.7 |
| MoVTeTaO HT(0.12) | 170±19 | 31±4 | 16 | 2.7 |
| MoVTeTaO HT(0.15) | 119±12 | 22±2 | 11 | 3.8 |
| MoVTeTaO MW | 174±21 | 40±3 | 20 | 2.6 |
| MoVTeTaO HT(P) | 136±12 | 21±2 | 11 | 3.3 |

$l_{mean}$: the mean of length of particles $d_{mean}$: the mean of diameter of *ab* planes; R: the radius of *ab* planes; C.*ab*.S: calculated *ab* planes surface area (m$^2$/g).

**Legends for Figures**

Figure S1. XRD patterns of MoVTeTa(Nb)O catalysts: (a) MoVTeNbO HT; (b) MoVTeTaO HT (0.09); (c) MoVTeTaO HT (0.12); (d) MoVTeTaO HT (0.15) (e) MoVTeTaO MW; (f) MoVTeTaO HT(P).

Figure S2. SEM images of MoVTeTaO catalysts: (a) MoVTeTaO HT (0.09); (b) MoVTeTaO HT (0.12); (c) MoVTeTaO HT (0.15) (d) MoVTeTaO MW; (e) MoVTeTaO HT(P); f) MoVTeTa SE [3].

Figure S3. (Left) The probability of finding Ta and Nb (calculated from DFT data); (Right) The HAADF-STEM images of M1-Ta HT(0.12) viewed down the [001] direction; a) at high magnification; b) center region of M1; and c) surface region of M1 particle [4].

Figure S4. Relative intensities of metal lattice sites in MoVTeTa(Nb)O M1 phases prepared by different synthesis methods. Relative intensities were obtained by normalizing the observed intensity by those of 100% Mo sites (S5, S6, S8, S10, and S11). The error bars were calculated by averaging multiple unit cells in multiple images.

Figure S5. a) 3D structural model of MoVTeTaO M1 phase oriented along [001] b) structural model of a 3 x 3 x 10 supercell of MoVTeTaO M1 phase oriented along [100]. These images were simulated by QSTEM software [5].

Figure S6. Experimentally observed maximum selectivity to acrylonitrile as a function of Ta occupancy in S9 of M1-Ta HT (0.09), (0.12), and (0.15), M1-Ta MW, M1-Ta HT(P), during propane ammoxidation; Reaction conditions: $C_3H_8$:$NH_3$:$O_2$:He=5.7:8.6:17.1:68.6 (%); 0.2 g catalyst; reaction temperature: 380-460°C.

Figure S7. The selectivity to acrylonitrile vs. Nb/Te ratios of $Mo_{0.6}V_{0.16-0.213}Nb_{0.055-0.10}Te_{0.10-0.18}O_x$ M1 phase catalysts [6].

Figure S8. Ta occupancy in S9 *versus* the Mo/V ratios in MoVTeTaO M1 catalysts.

Figure S9. Ta occupancy in S9 *versus* the Mo/V ratios in the catalytic center (S2-S4-S4-S7-S7) of MoVTeTaO M1 phase.

Figure S10. Combined yield of propylene and acrylonitrile, $Y(C_3H_6+ACN)$, as a function of probability of 1-2 $V^{5+}$ in the newly proposed catalytic center (S3-S4-S4-S7-S7) of M1-Ta HT(0.09), M1-Ta HT(0.12), M1-Ta HT(0.15), M1-Ta MW, and HT(P) in propane ammoxidation; Reaction conditions: $C_3H_8$:$NH_3$:$O_2$:He=5.7:8.6:17.1:68.6 (%); total flow rate, 26.3 mL•min$^{-1}$; 0.2 g catalyst; reaction temperature: 420°C.

Figure S11. Arrhenius plots of k" of propane consumption over Ta HT(0.09), M1-Ta HT(0.12), M1-Ta HT(0.15), M1-Ta MW, and HT(P) catalyst during propane ammoxidation; Reaction conditions: $C_3H_8$:$NH_3$:$O_2$:He=5.7:8.6:17.1:68.6; total flow rate, 26.3 mL•min$^{-1}$; 0.2 g catalyst; reaction temperature: 350-440°C.

Figure S12. Reaction rate constant of propane consumption, k", vs. total V content in the catalytic center (S2-S4-S4-S7-S7) of M1-Ta HT(0.09), M1-Ta HT(0.12), M1-Ta HT(0.15), M1-Ta MW, and HT(P) in propane ammoxidation; Reaction conditions: $C_3H_8$:$NH_3$:$O_2$:He=5.7:8.6:17.1:68.6; total flow rate, 26.3 mL•$min^{-1}$; 0.2 g catalyst; reaction temperature: 420°C.

Figure S13. Reaction rate constant, k", based on BET surface areas plotted as a function of probability of V>2 in the catalytic center (S2-S4-S4-S7-S7) of M1-Ta HT(0.09), M1-Ta HT(0.12), M1-Ta HT(0.15), M1-Ta MW, and HT(P) in propane ammoxidation; Reaction conditions: $C_3H_8$:$NH_3$:$O_2$:He=5.7:8.6:17.1:68.6;total flow rate, 26.3 mL•$min^{-1}$; 0.2 g catalyst; reaction temperature: 420°C.

Figure S1. TIF

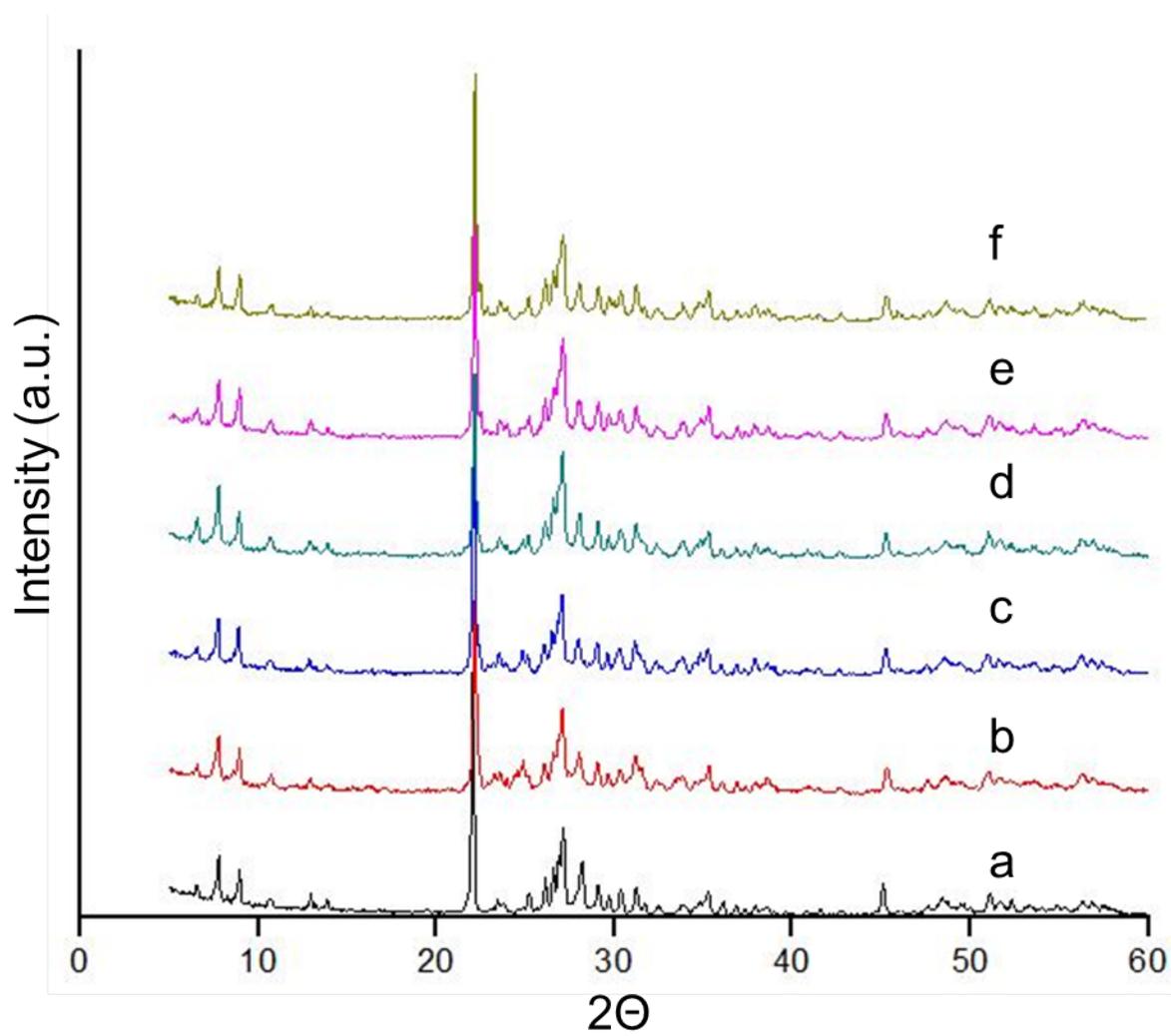

Figure S1. TIF

Figure S2. TIF

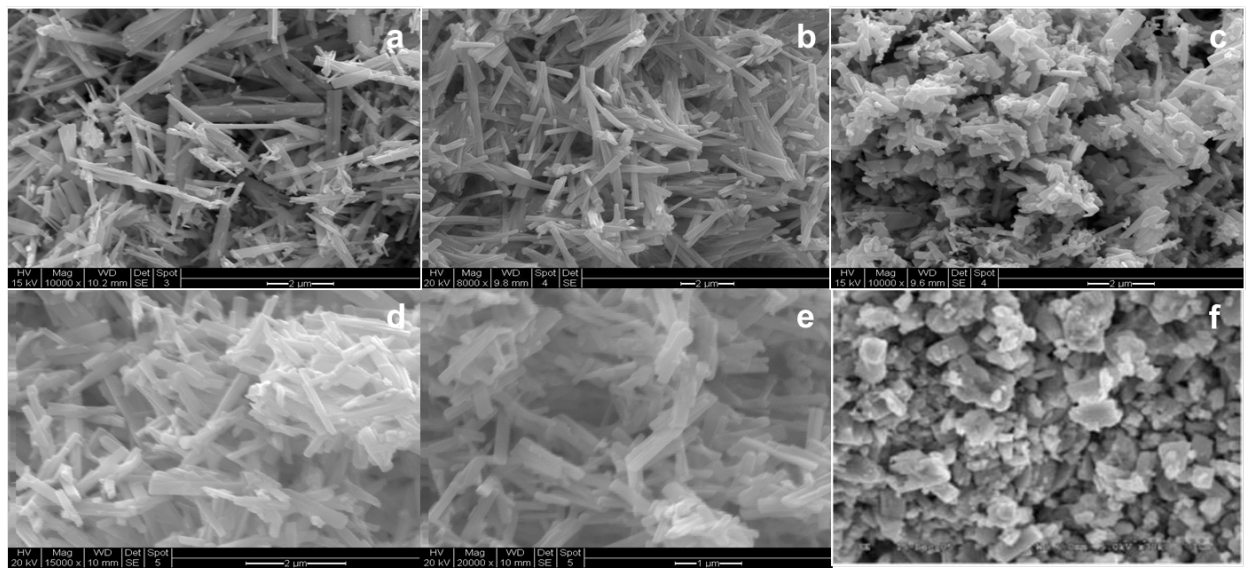

Figure S3. TIF

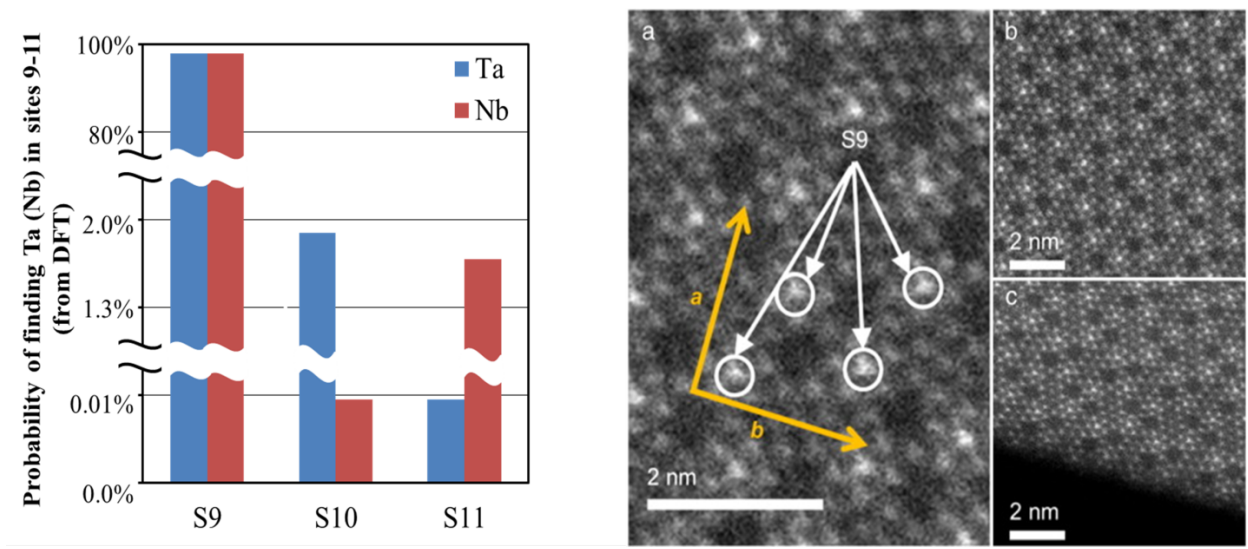

Figure S4. As is

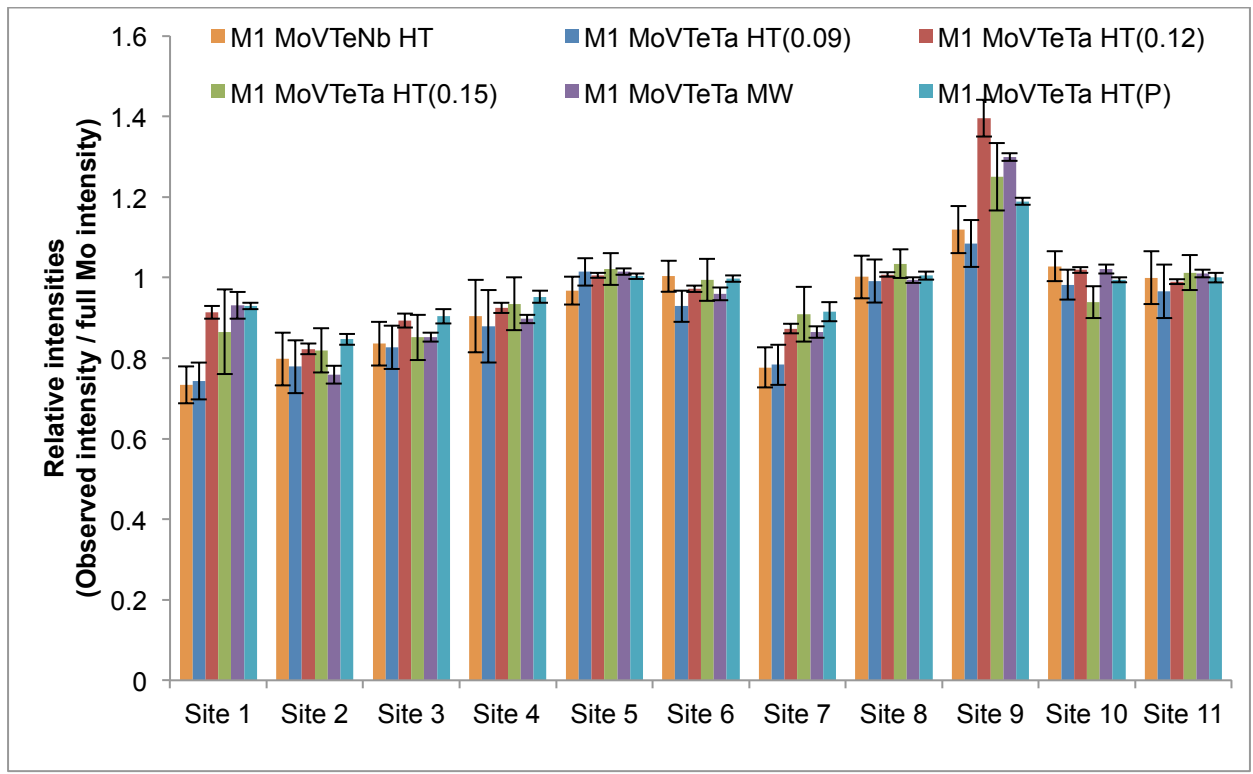

Figure S5. TIF

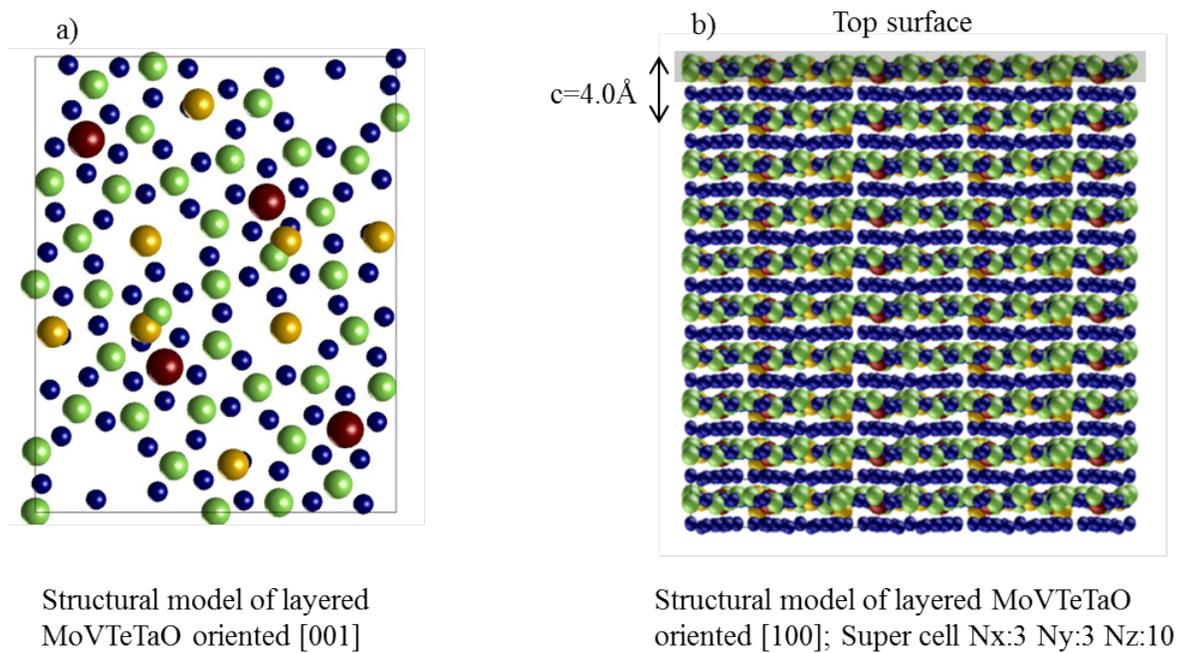

a) Structural model of layered MoVTeTaO oriented [001]

b) Structural model of layered MoVTeTaO oriented [100]; Super cell Nx:3 Ny:3 Nz:10

Figure S6. As is

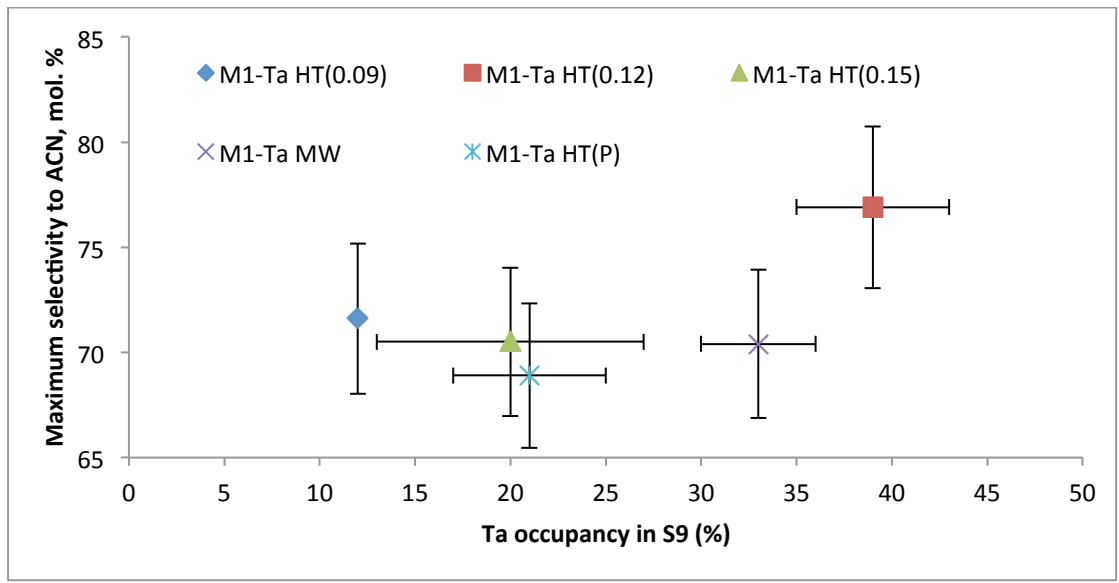

Figure S7. TIF

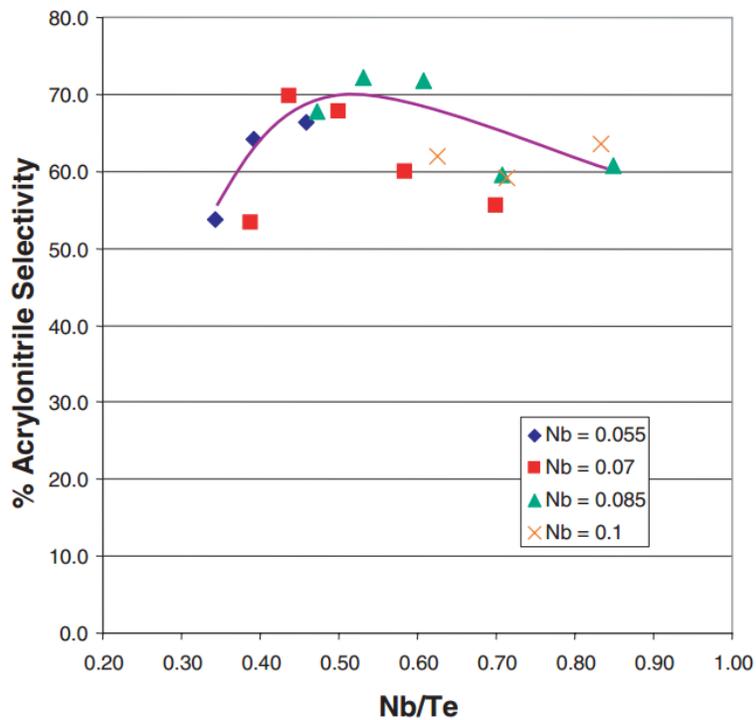

Figure S8. As is

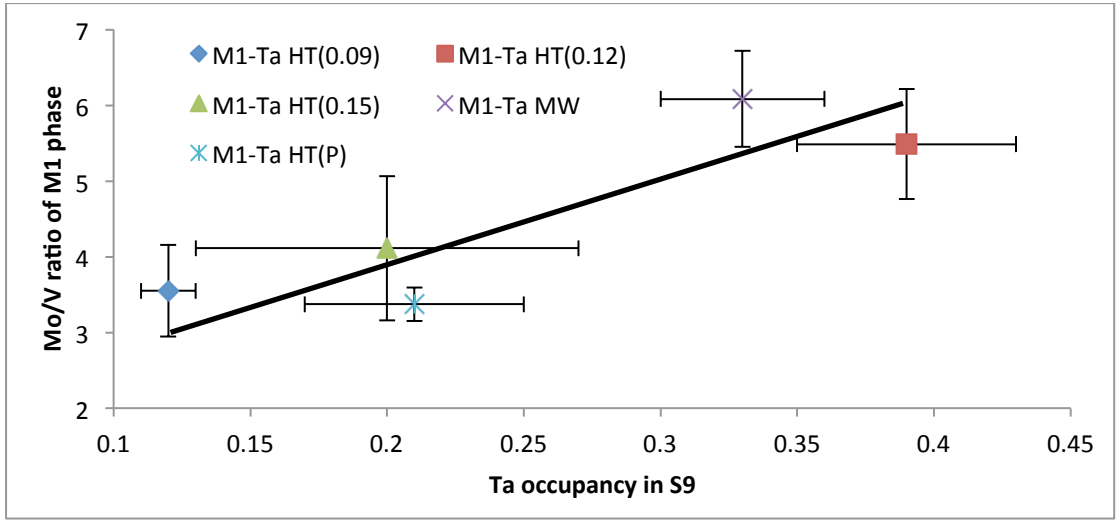

Figure S9. As is

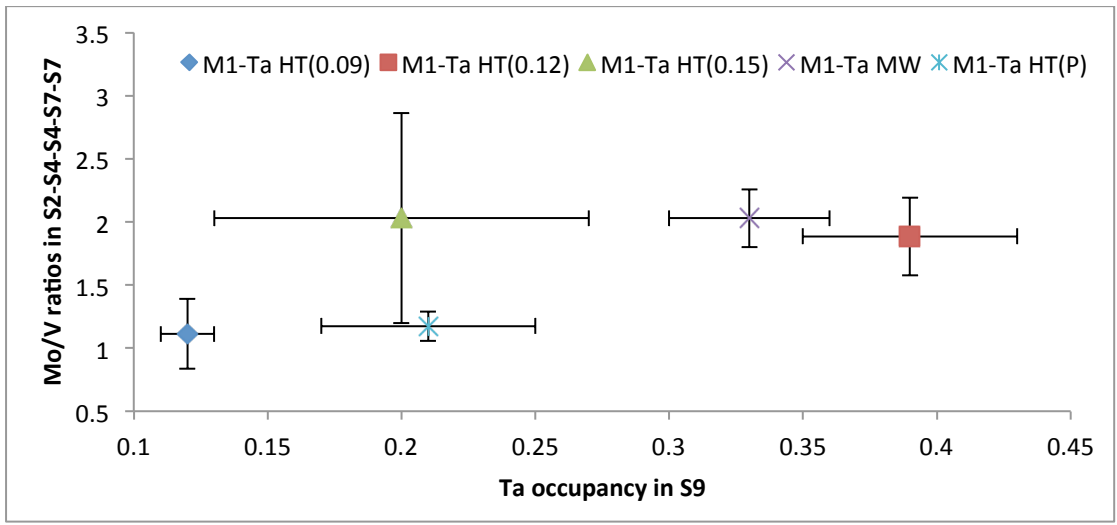

Figure S10. As is

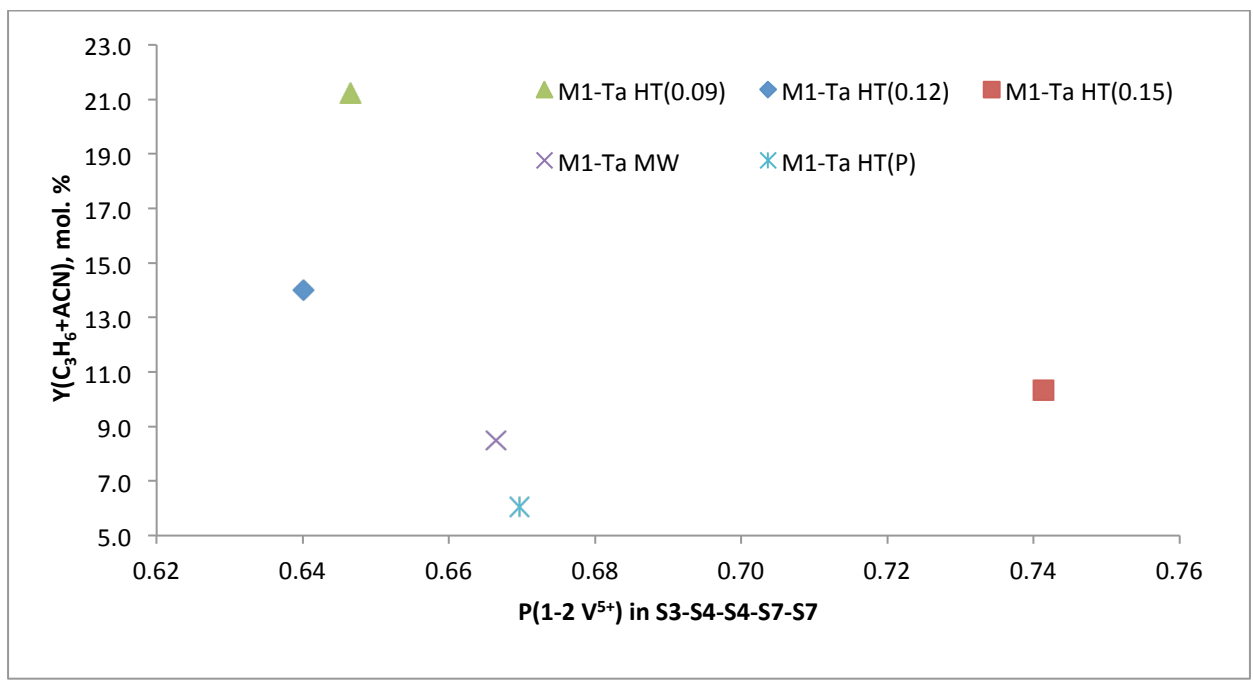

Figure S11. As is

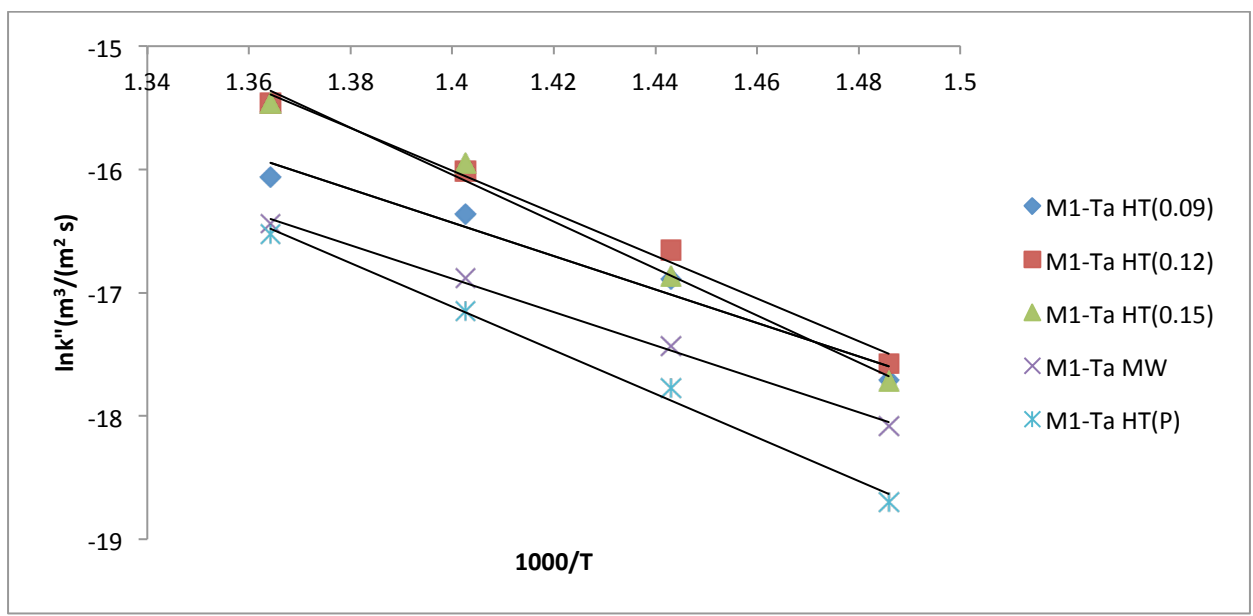

Figure S12. As is

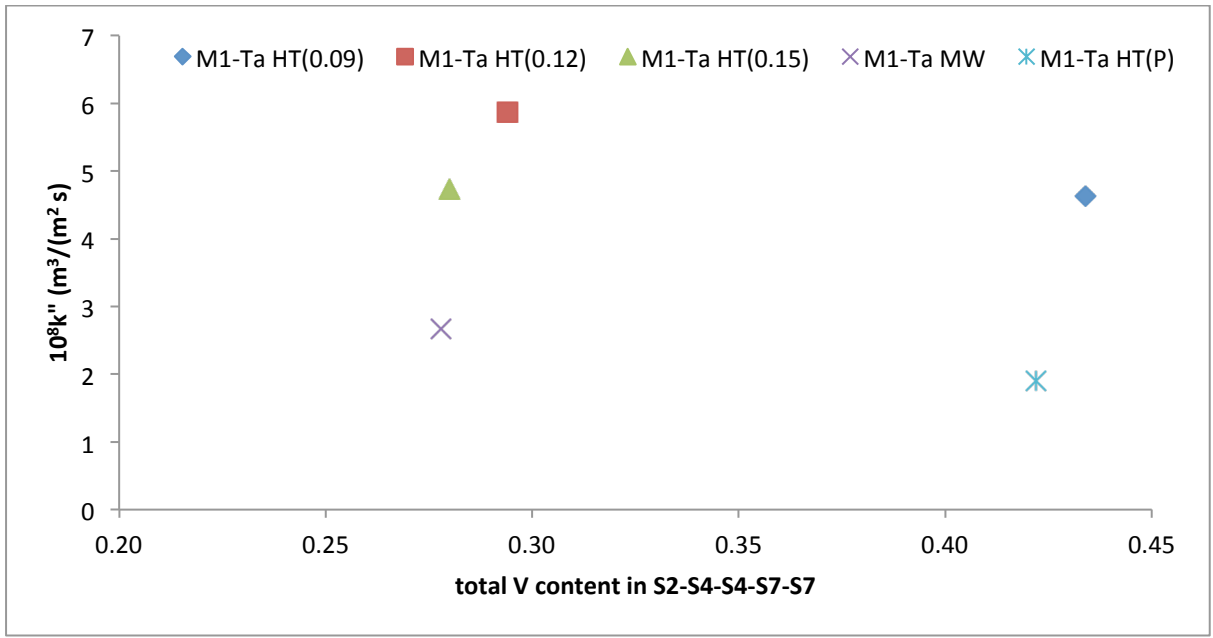

Figure S13. As is

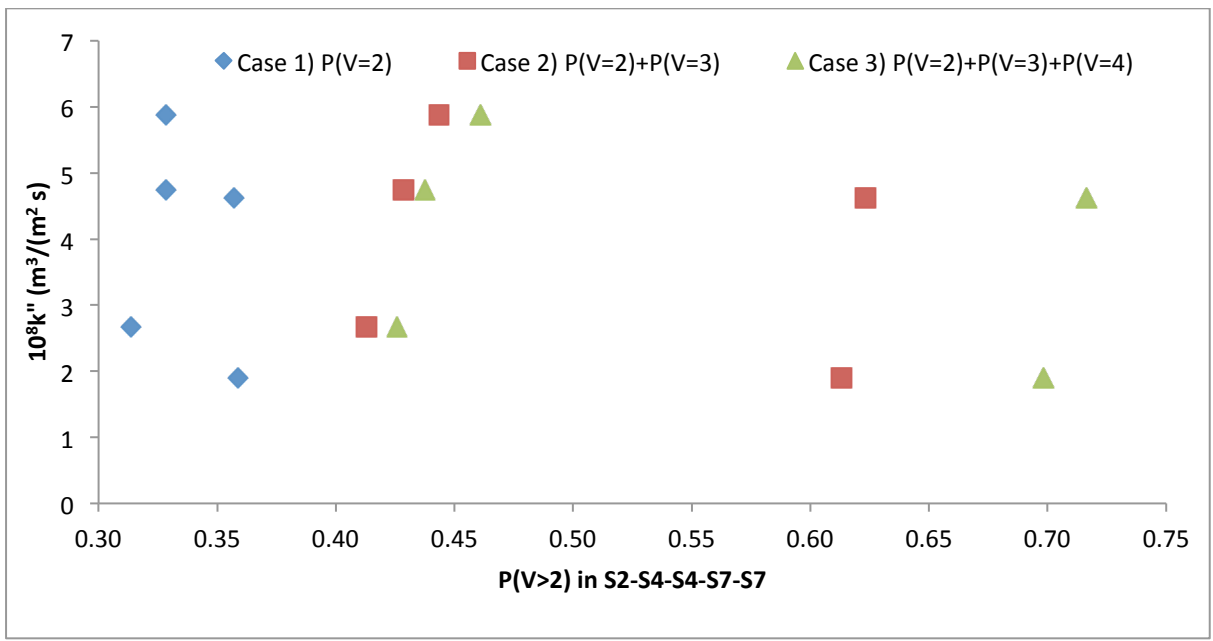

The 1$^{st}$ order irreversible reaction rate constant [7] was calculated as follows:

A → products, for any constant $\varepsilon_A$

$-r''_A = -\frac{1}{S}\frac{dN_A}{dt} = k''C_A$, $[\frac{mol\ reacted}{m^2 cat.surf.\cdot s}]$

$k''\tau'' = -(1+\varepsilon_A) \ln(1-X_A) - \varepsilon_A X_A$

where S is catalyst surface; $\tau''$ is the catalyst area-time time; $\varepsilon_A$ is the fractional change in volume of the system for complete conversion of reactant A; $X_A$ is a fraction of A converted.